\documentclass[10pt]{article}
 
\usepackage{amsmath,amsthm}
\usepackage{amsthm}
\usepackage{amssymb}
\usepackage{amsfonts} 
\usepackage[latin1]{inputenc}
\usepackage{dsfont}
\usepackage{nicefrac}
\usepackage{graphicx}
  \usepackage{stmaryrd}
\usepackage{pgfplots}

\usepackage[colorlinks=true,citecolor = purple,urlcolor=purple, linkcolor=blue,
filecolor=magenta]{hyperref}

\usepackage{fullpage,amsmath,amsfonts,amssymb,amsthm}
\usepackage{arydshln}
\usepackage{cite}
\usepackage{braket}

\usepackage[font=small,labelfont=bf]{caption} 

\newtheorem{Definition}{Definition}

\newtheorem{remark}{Remark}

\def\be{\begin{equation}}
\def\ee{\end{equation}}
\def\bea{\begin{eqnarray}}
\def\eea{\end{eqnarray}}

\newcommand\Om\Omega

 \def\1{\'{\i}}

\numberwithin{equation}{section}

\begin{document}

\thispagestyle{empty}
\begin{center}
\Large{{\bf Construction of polynomial algebras from intermediate Casimir invariants of Lie algebras}}
\end{center}
\vskip 0.5cm
\begin{center}
\textsc{Danilo Latini, Ian Marquette and Yao-Zhong Zhang}
\end{center}
\begin{center}

School of Mathematics and Physics, The University of Queensland, Brisbane, QLD 4072, Australia
\end{center}
\begin{center}
	\footnotesize{\textsf{d.latini@uq.edu.au}, \textsf{i.marquette@uq.edu.au}, \textsf{yzz@maths.uq.edu.au}}
\end{center}
\vskip 0.5cm

\vskip  1cm
\hrule

  \begin{abstract}
 	\noindent  We propose a systematic procedure to construct polynomial algebras from intermediate Casimir invariants arising from (semisimple or non-semisimple)  Lie algebras $\mathfrak{g}$. In this approach, we deal with explicit polynomials in the enveloping algebra of $\mathfrak{g} \oplus \mathfrak{g} \oplus \mathfrak{g}$. We present explicit examples to show how these Lie algebras can display different behaviours and can lead to Abelian algebras, quadratic algebras or more complex structures involving higher order nested commutators.  Within this framework, we also demonstrate how virtual copies of the Levi factor of a Levi decomposable Lie algebra can be used as a tool to construct \textquotedblleft copies\textquotedblright of polynomial algebras. Different schemes to obtain polynomial algebras associated to algebraic Hamiltonians have been proposed in the literature,  among them the use of commutants of various type. The present approach is different and relies on the construction of intermediate Casimir  invariants in the enveloping algebra $\mathcal{U}(\mathfrak{g} \oplus \mathfrak{g} \oplus \mathfrak{g})$.
 \end{abstract}
\vskip 0.35cm
\hrule

%
%
%
%
%
\section{Introduction}
\label{intro}

Finite-dimensional integrable and superintegrable systems represent special subclasses of dynamical Hamiltonian systems. From the algebraic point of view, one of their main differences lies in the algebraic structures generated by the integrals of motion. In the integrable case, all integrals are in involution and as such only abelian algebras arise. In the more interesting superintegrable case, the presence of additional constants of motion implies the existence of non-abelian algebras spanned by the integrals \cite{mil13}.
Usually, these non-abelian structures are Lie, quadratic, cubic or higher order polynomial algebras \cite{Bonatsos:1993st, Bonatsos:1994fi, doi:10.1063/1.2399359, doi:10.1063/1.3013804, doi:10.1063/1.3096708, Marquette_2010, Marquette_2011, Post2011, doi:10.1063/1.4816086, 1751-8121-47-20-205203, Marquette2019a, Marquette2019, latini2022polynomial}. The standard example that is often recalled in the literature is the $3$D Kepler-Coulomb system. For this maximally superintegrable (MS) model the symmetry algebra, generated by the angular momentum and  Laplace-Runge-Lenz \cite{goldstein2002classical, doi:10.1119/1.9745, doi:10.1119/1.10202} vectors is a quadratic algebra involving the Hamiltonian as its central element. Then, in the elliptic case, when we restrict to the subspace of constant negative energy, it linearizes to the Lie algebra $\mathfrak{so}(4) \cong \mathfrak{su}(2) \oplus \mathfrak{su}(2)$. Another standard example is the $3$D isotropic harmonic oscillator, for which the symmetry algebra is known to be the eight-dimensional Lie algebra $\mathfrak{su}(3)$, generated by suitable combinations of the angular momentum vector and Demkov-Fradkin tensor \cite{Dem, doi:10.1119/1.1971373} components.
Both of these models are of course maximally superintegrable and are in fact the only two Euclidean MS systems among the family of radially symmetric ones. This is a consequence of Bertrand's Theorem \cite{Bertrand1873}.

In a recent paper \cite{latini2020embedding}, we have shown that a subclass of  superintegrable models (in any dimension $n$) generalizing spherically symmetric ones are in fact characterized by symmetry algebras satisfying the relations of the \emph{generalized Racah algebra} $R(n)$ (for a comprehensive review about $R(n)$ we refer the reader to the recent works \cite{bie2020racah, https://doi.org/10.48550/arxiv.2105.01086}). Among the models in this class are the well-known Smorodinsky-Winternitz \cite{FRIS1965354, Makarov1967, EVANS1990483, doi:10.1063/1.529449, 2021correa} and generalized Kepler-Coulomb (also known as Evans-Verrier) \cite{doi:10.1063/1.2840465, 2011SIGMA7054T, 175181214224245203, 17518121468085206} systems. To obtain these results we relied on the so-called left and right partial Casimir invariants, commonly encountered in the framework of \emph{coalgebra symmetry approach to superintegrability}  \cite{Ballesteros1996, 0305-4470-31-16-009, 1742-6596-175-1-012004}, that can be constructed from suitable linear combinations of the generalized Racah $R(n)$ generators \cite{lat21}.  The generalized Racah algebra $R(n)$ previously appeared as the symmetry algebra of the generic superintegrable model on the $(n-1)$-sphere (see \cite{bie2020racah}  and references therein) and pseudo-sphere \cite{Kuru2020}. It was proposed in \cite{Debie2017}  as a higher-rank generalisation of the rank one Racah algebra $R(3)$, which is in turn the symmetry algebra of the generic superintegrable model on the 2-sphere \cite{ Kalnins_2007, 2011SIGMA7051K}.

In light of the equivalence between the superintegrable system on the $2$-sphere and the Racah problem for $\mathfrak{su}(1,1)$ \cite{gen14, Genest2014} (see also \cite{post15} for the equivalent problem on the $3$-sphere), the quantum integrals have been reinterpreted as those operators that realize the so-called \emph{intermediate Casimir} elements arising in the addition of three irreducible representations of $\mathfrak{su}(1,1)$, whereas the Hamiltonian arise when an operator realization of the \emph{full Casimir} element is considered. In this setting, the Racah algebra $R(3)$ is understood as the subalgebra of $\mathcal{U}(\mathfrak{su}(1,1)) \otimes \mathcal{U}(\mathfrak{su}(1,1))  \otimes \mathcal{U}(\mathfrak{su}(1,1))$ generated by these Casimir elements. From the point of view of superintegrability, the crucial result relies on the fact that from this model all second-order $2$D conformally flat superintegrable systems \cite{Kalnins05,Kalnins052,  Kalnins053} can be obtained as limiting cases \cite{Kalnins_2013}.  This result has also been extended to dimension three \cite{Capel_2015}. This fact highlights the fundamental importance that the quadratic algebra $R(3)$ has in the theory of superintegrable systems. Within this framework, symmetry algebras of superintegrable systems are understood as the Poisson/commutator algebras generated by the classical/quantum integrals of motion associated to a given Hamiltonian function/operator. The integrals are polynomials in the momenta/finite-order partial differential operators depending on the context one is working on, i.e. classical or quantum mechanics. 

Recently, different schemes to obtain finitely generated quadratic algebras associated to superintegrable systems from a purely algebraic point of view have been proposed \cite{CAMPOAMORSTURSBERG2021168378, cor21, CAMPOAMORSTURSBERG2022168694, cam2022}. These approaches rely on enveloping algebras of Lie algebras. In this framework, the polynomial symmetry algebras are understood as the subspaces in the enveloping algebra that commute with a given algebraic Hamiltonian.  The algebraic Hamiltonian and integrals of motion are viewed as polynomials in the enveloping algebra of a given Lie algebra. Thus, polynomial algebraic structures in this setting turn out to be independent of the realizations. This led to abstract symmetry algebras that are shared by \emph{formally equivalent} systems obtained when different realizations of the generators in terms of differential operators are considered. Along this more abstract line of research, in this paper we aim to obtain polynomial symmetry algebras generated by intermediate Casimir invariants constructed  from a Lie algebra $\mathfrak{g}$, not necessarily semisimple, in a purely algebraic way.  This will allow us to consider the associated algebraic Hamiltonians which will then have algebraic integrals of motion.  Integrability and superintegrability will then occur depending on explicit choices of the realizations.

\noindent The paper is organized as follows: In Section \ref{sec2}, after recalling some general notions and definitions related to polynomial Casimir invariants of Lie algebras that will be useful throughout the paper, we introduce a systematic procedure to construct polynomial symmetry algebras from intermediate Casimir invariants of Lie algebras. The connection with the Racah algebra $R(3)$ is made explicit in Section \ref{sec3}, where the construction is initially applied to the simple three dimensional Lie algebra $\mathfrak{sl}_2(\mathbb{F})$ and then to the six dimensional Lie algebra $\mathfrak{so}(1,3)$. Then,  taking into account the classification results reported in \cite{sno14} for indecomposable Lie algebras up to dimension six, in Sections \ref{sec4}, \ref{sec5} and \ref{sec6} we present several explicit examples of polynomial algebras constructed from some representatives of \emph{nilpotent}, \emph{solvable} and \emph{Levi decomposable} Lie algebras, respectively. This allows us to shed some light on the different algebraic structures obtained in this way and, due to the specific relations encountered, to propose a proper definition for the notion of closure that, for some cases, necessitates the introduction of higher order nested commutators.  For some specific Abelian case, we show how the use of subalgebras can still provide an alternative way to obtain different polynomial structures. Also, taking as an example the six dimensional Levi decomposable Lie algebra $\mathfrak{sl}_2 \niplus \mathfrak{n}_{3,1}$, we discuss on how the virtual copies of the Levi factor \cite{cam09} can be used to generate what we might define to be a virtual copy of the Racah algebra $R(3)$.
Finally, Section \ref{sec7} is devoted to some concluding remarks.

\section{Polynomial algebras from intermediate Casimir invariants}
\label{sec2}

Let us consider a $d$-dimensional Lie algebra $\mathfrak{g} := \{X_1, \dots, X_d \, : \, [X_i, X_j] = C^k_{i j} X_k\}$ over a field $\mathbb{F}$.
\begin{Definition}
A Casimir invariant of $\mathfrak{g}$ is an element $C \in \mathcal{U}(\mathfrak{g})$ which commutes with all generators of $\mathfrak{g}$, i.e. $[X_{i},C]=0 \quad \forall \, i=1, \dots, d$. 
\end{Definition}
\noindent We recall that any element of $\mathcal{U}(\mathfrak{g})$ is a linear combination of (ordered) basis elements:
\begin{equation}
\{X_1^{a_1}  X_2^{a_2}  \dots X_d^{a_d}  \, :\,  a_1, a_2, \dots, a_d \in \mathbb{Z}_{\geq 0} \} \, .
\label{eq:basis}
\end{equation}
Multiplication in $\mathcal{U}(\mathfrak{g})$ is associative (it glues monomials), and the ordering is kept by endowing $\mathcal{U}(\mathfrak{g})$ with the additional relation $X_i X_j-X_j X_i = C_{ij}^k X_k \equiv [X_i, X_j]$. For example: 
$$X_j X_i X_k  \to  (X_i X_j -C_{ij}^\ell X_\ell)X_k = X_i X_j X_k + \text{l.o.t.} \qquad (i \leq j \leq k) \, .$$

\noindent That said, we indicate a (polynomial) Casimir invariant $C \in \mathcal{U}(\mathfrak{g})$ of order $p$ as:
\begin{equation}
	C(X_1,\dots,X_d) = \sum_{\substack{a_1,\dots, a_d \\ a_1+\dots+ a_d \leq p}} f_{a_1 \dots a_d}X_1^{a_1} \dots X_d^{a_d} \, , \qquad  f_{a_1 \dots a_d} \in \mathbb{F}.
\label{cas}
\end{equation}

\noindent We restrict our analysis to polynomial Casimir invariants. However, we emphasise that in general, for non-semisimple Lie algebras, rational or trascendental Casimir invariants can exist $($they are referred to as generalised Casimir invariants$)$ \cite{sno14, strursberg2018group}.

\noindent Given a non-semisimple Lie algebra $\mathfrak{g}$, the search of its Casimir invariants is a difficult task.  Contrary to semisimple Lie algebras, for which Casimir invariants can be constructed directly,  for non-semisimple Lie algebras this is no longer the case and different approaches have been developed trying to solve this problem at least for certain classes (see e.g. \cite{BELTRAMETTI1966, Quesne1988, cam05, strursberg2018group, sno14, cam09, Alshammari2018}). Among those, an analytical method that is often considered is the one in which the Lie algebra generators are realized in the space  $C^\infty(\mathfrak{g}^*)$ as the differential operators:
\begin{equation}
\hat{X}_i = \sum_{j,k=1}^dC_{ij}^k x_k \partial_{x_j} \, ,
\label{eq:diffreal}
\end{equation}  
where $(x_1, \dots, x_d)$ represent coordinates in the dual vector space $\mathfrak{g}^*$ of $\mathfrak{g}$.  In this approach, the problem restricts in finding the solutions of the following system composed by $d:=\text{dim}(\mathfrak{g})$ linear PDEs:
\begin{equation}
\begin{cases}
 \sum_{j,k=1}^dC_{1j}^k x_k \partial_{x_j} F(x_1, \dots, x_d)&=0 \\
 \sum_{j,k=1}^dC_{2j}^k x_k \partial_{x_j} F(x_1, \dots, x_d)&=0\\
& \vdots \\
\sum_{j,k=1}^d C_{dj}^k x_k \partial_{x_j} F(x_1, \dots, x_d)&=0 \, .
\end{cases}
\label{eq:casinv}
\end{equation}

\noindent The maximal number  $N(\mathfrak g)$ of functionally independent solutions of \eqref{eq:casinv} can be found from the following formula \cite{BELTRAMETTI1966,pau66}:
 \begin{equation}
N(\mathfrak g)= \text{dim} (\mathfrak g)- \text{rank} ||C_{ij}^k x_k|| \, \quad (i,j,k = 1, \dots, \text{dim}(\mathfrak{g})) \label{bb} \, ,
\end{equation}
where $C_{ij}^k$ are the structure constants and the maximum rank of the
$\text{dim}(\mathfrak{g}) \times \text{dim}(\mathfrak{g})$ matrix $||C_{ij}^k x_k||$ is computed by considering $x_k$ as independent variables.
Then, if one restricts to consider just polynomial solutions of \eqref{eq:casinv},  say $F^{(r)}(x_1,\dots, x_d) \equiv P^{(r)}(x_1, \dots, x_d)$ for $r=1,\dots, N_p(\mathfrak{g})$, $N_p(\mathfrak{g})$ being the number of polynomial solutions, then the polynomial Casimir invariants in $\mathcal{U}(\mathfrak{g})$ are obtained as a result of a symmetrisation procedure of the monomials appearing in the polynomials $P^{(r)}(x_1,\dots,x_d)$ after replacing $x_i$ with $X_i$. 

\noindent Formally, we might write $P^{(r)}(x_1,\dots, x_d) \to \text{Sym}(P^{(r)})(X_1,\dots, X_d)$, where $\text{Sym}$ is the symmetrization map acting on monomials as:
\begin{equation}
\text{Sym}: \quad x_{i_1}\dots x_{i_k} \quad \to\quad  \frac{1}{k!}\sum_{\sigma \in \Sigma_k} x_{\sigma(i_1)}\dots {x_{\sigma(i_k)}} \, ,
\label{smap}
\end{equation}
\noindent $\Sigma_k$ being the symmetric group in $k$ letters \cite{strursberg2018group}. Just as an illustrative example let us focus on the simple Lie algebra $\mathfrak{sl}_2(\mathbb{F})$ in the basis $\{X_1, X_2, X_3\}$ with commutation table \cite{sno14}:
\begin{center}
	\begin{tabular}{| l | c|c | r| }
		\hline
		& $X_1$ & $X_2$ & $X_3$ \\ \hline
		$X_1$ & 0 & 2$X_1$ &$-X_2$ \\ \hline
		$X_2$ & $-2X_1$ & 0 & $2X_3$\\ \hline
		$X_3$ & $X_2$ & $-2X_3$ & 0\\
		\hline
	\end{tabular}
\end{center}

\noindent Of course, in this case one has $N(\mathfrak g)=1$ and the vector fields \eqref{eq:diffreal} turn out to be:
\begin{equation}
\hat{X}_1=2x_1\partial_{x_2}-x_2 \partial_{x_3} \, ,\qquad
\hat{X}_2=-2x_1\partial_{x_1}+ 2x_3 \partial_{x_3} \, ,\qquad
\hat{X}_3=x_2\partial_{x_1}- 2x_3 \partial_{x_2} \, .
\label{eq:vecfields} 
\end{equation}
 Then, to obtain the Casimir operator we need to find the solution of the following system of PDEs:
\begin{equation}
 	\begin{cases}
 	\hat{X_1}F&=2x_1 F_{x_2}-x_2 F_{x_3}=0
 	\\
 	\hat{X_2}F&=-2x_1 F_{x_1}+2x_3 F_{x_3}=0
 	\\
 	\hat{X_3}F&=x_2 F_{x_1}-2x_3 F_{x_2}=0 \, ,
 	\end{cases}
 	\end{equation}
 \noindent  which is solved for:
 \begin{equation}
 F(x_1,x_2,x_3)= x_2^2+4 x_1x_3 \, .
 \end{equation}
At this point, by symmetrising the result we get the following quadratic element in $\mathcal{U}(\mathfrak{sl}_2(\mathbb{F}))$:
 \begin{equation}
C(X_1,X_2,X_3)= X_2^2+2 (X_1X_3+X_3X_1)=X_2^2+4  X_1X_3+2X_2  \, ,
\end{equation}
\noindent which commutes with all the basis generators $X_i \in \mathfrak{sl}_2(\mathbb{F})$:
\begin{equation}
[X_1,C]=[X_2,C]=[X_3,C]=0 \, ,
\label{commcas}
\end{equation}
as it can be straightforwardly checked by computing explicitly these commutators.

\subsection{The general construction}
\label{sec2.1}
In this Section \ref{sec2.1} we propose a systematic procedure to construct polynomial algebras from intermediate Casimir invariants of Lie algebras. To this aim, let $\mathfrak{g} = \bigl(X_1, \dots, X_d\,  : 	\,   [X_i,X_j]=C_{ij}^k X_k\bigl)$ be a  $d$-dimensional Lie algebra over a field $\mathbb{F}$ endowed with $r = 1,2, \dots, N_p(\mathfrak{g})$ functionally independent (non-linear) polynomial Casimir elements, which we denote as:
\begin{equation}
C^{(r)}(X_1,\dots,X_d) =  \sum_{\substack{a_1,\dots, a_d \\ a_1+\dots+ a_d \leq p^{(r)}}} f^{(r)}_{a_1 \dots a_d} X_1^{a_1} \dots X_d^{a_d} \, .
\label{asi}
\end{equation}
\noindent Consider three copies of $\mathfrak{g}$ with basis elements $\{X_i^{[\alpha]}\}$, $i=1, \dots, d$, $\alpha=1,2,3$ such as:
\begin{align}
&[X_i^{[1]}, X_j^{[1]}] = C_{ij}^k X_k^{[1]} \quad [X_i^{[2]}, X_j^{[2]}] = C_{ij}^k X_k^{[2]} \quad [X_i^{[3]}, X_j^{[3]}] = C_{ij}^k X_k^{[3]}  \label{eq11}\\
&[X_i^{[1]}, X_j^{[2]}] = 0 \hskip 1.3cm [X_i^{[2]}, X_j^{[3]}] = 0 \hskip 1.3 cm [X_i^{[3]}, X_j^{[1]}] =0 \label{eq12}  \, .
\end{align}
\noindent Each copy of the algebra is then endowed with the polynomial Casimir invariants:
\begin{equation}
C_\alpha^{(r)}(X_1^{[\alpha]},\dots,X_d^{[\alpha]}) =  \sum_{\substack{a_1,\dots, a_d \\ a_1+\dots+ a_d \leq p^{(r)}}} \hskip -0.3cm f^{(r)}_{a_1 \dots a_d}\bigl(X_1^{[\alpha]}\bigl)^{a_1} \dots \bigl(X_d^{[\alpha]}\bigl)^{a_d} \qquad (r=1, \dots, N_p(\mathfrak{g})) \, .
\label{asic}
\end{equation}
\noindent At this point,  considering two copies $(\alpha,\beta)$ we can construct the new elements:
\begin{equation}
 X_i^{[1, 2]}:=X_i^{[1]} + X_i^{[2]}  \, , \quad  X_i^{[2,3]}:=X_i^{[2]} + X_i^{[3]} \, , \quad  X_i^{[1,3]}:=X_i^{[1]} + X_i^{[3]} \, \qquad (i=1, \dots, d) \, ,
\label{addtwo} 
\end{equation}
which satisfy by construction the commutation relations:
\begin{align}
[X_i^{[1,2]}, X_j^{[1,2]}] = C_{ij}^k X_k^{[1,2]} \quad [X_i^{[2,3]}, X_j^{[2,3]}] = C_{ij}^k X_k^{[2,3]} \quad [X_i^{[1,3]}, X_j^{[1,3]}] = C_{ij}^k X_k^{[1,3]} \, ,  
\end{align}
\noindent as it can be directly checked:
\begin{align*}
[X_i^{[\alpha,\beta]}, X_j^{[\alpha,\beta]}]=[X_i^{[\alpha]} + X_i^{[\beta]} , X_j^{[\alpha]} + X_j^{[\beta]} ] =[X_i^{[\alpha]}, X_j^{[\alpha]}]+ [X_i^{[\beta]}, X_j^{[\beta]}]&=C_{ij}^kX_k^{[\alpha]}+C_{ij}^kX_k^{[\beta]}\\&=C_{ij}^k X_k^{[\alpha, \beta]} \, .
\end{align*}
\noindent Each copy  $(\alpha, \beta)$ of the algebra is then endowed with the \emph{intermediate} (polynomial) Casimir invariants:

\begin{equation}
C_{\alpha \beta}^{(r)}(X_1^{[\alpha,\beta]},\dots,X_d^{[\alpha, \beta]}) =  \sum_{\substack{a_1,\dots, a_d \\ a_1+\dots+ a_d \leq p^{(r)}}} \hskip -0.3cm f^{(r)}_{a_1 \dots a_d}\bigl(X_1^{[\alpha, \beta]}\bigl)^{a_1} \dots \bigl(X_d^{[\alpha, \beta]}\bigl)^{a_d}  \qquad (r=1, \dots, N_p(\mathfrak{g})) \, ,
\label{eq:casinvij}
\end{equation}
\noindent that are symmetric by construction $C_{\beta \alpha}^{(r)}=C^{(r)}_{\alpha \beta}$. Then, if we consider the total number of copies $(1, 2, 3)$ we can introduce the elements:
\begin{equation}
X_i^{[1,2,3]}:=X_i^{[1]}+X_i^{[2]}+X_i^{[3]} \qquad (i=1, \dots, d) \, .
\label{total}
\end{equation}
\noindent Clearly, these elements satisfy again the commutation relations:
\begin{align}
&[X_i^{[1,2,3]}, X_j^{[1,2,3]}] = C_{ij}^k X_k^{[1,2,3]} \, .
\end{align}
\noindent To this algebra are associated the polynomial Casimir invariants:
\begin{equation}
{\small C_{123}^{(r)}\bigl(X_1^{[1,2,3]},\dots,X_d^{[1,2,3]}\bigl)\, =  \sum_{\substack{a_1,\dots, a_d \\ a_1+\dots+ a_d \leq p^{(r)}}} \hskip -0.3cm f^{(r)}_{a_1 \dots a_d}\bigl(X_1^{[1,2,3]}\bigl)^{a_1} \dots \bigl(X_d^{[1,2,3]}\bigl)^{a_d}  \, \qquad (r=1, \dots, N_p(\mathfrak{g})) }\, .
\label{eq:totalcas}
\end{equation}
With this procedure, the following  $N_p(\mathfrak{g})$ sets composed by the one, two and three indices Casimir invariants in $\mathcal{U}(\mathfrak{g} \oplus \mathfrak{g} \oplus \mathfrak{g} ) \cong \mathcal{U}(\mathfrak{g}) \otimes \mathcal{U}(\mathfrak{g}) \otimes \mathcal{U}(\mathfrak{g})$ can be defined:
\begin{equation}
\mathcal{C}^{(r)}:=\{C_\alpha^{(r)}, C_{\alpha \beta}^{(r)}, C_{123}^{(r)}\} \qquad (r=1, \dots, N_p(\mathfrak{g})) \, .
\label{totalset}
\end{equation}
\noindent By construction, these Casimir invariants commute with the elements $X_i^{[1,2,3]}$, i.e.:
\begin{equation}
[\mathcal{C}^{(r)}, X_i^{[1,2,3]}]=0 \, , \qquad \forall \, i=1, \dots, d \, , \, \quad r=1, \dots, N_p(\mathfrak{g}) \, .
\label{eq:gens}
\end{equation}
\begin{remark}
The construction in terms of direct sums could be also seen as a way to deal with abstract tensor products from a computational point of view.  These elements, in fact, could be understood as the Casimir invariants arising from the application of the primitive coproduct map:
\begin{equation}
\Delta: \mathcal{U}(\mathfrak{g})\to \mathcal{U}(\mathfrak{g}) \otimes \mathcal{U}(\mathfrak{g}) \, .
\label{map}
\end{equation}
This map satisfies the properties \cite{doi:10.1142/10898}:
\begin{equation}
\Delta(U \dot V)=\Delta(U) \Delta(V) \qquad \Delta(a_0 U+b_0 V)=a_0 \Delta(U)+b_0 \Delta(V) \qquad \forall \,U,V \in \mathcal{U}(\mathfrak{g}) \, , \, \forall \,a_0,b_0 \in \mathbb{F} \, .
\label{eq:homom}
\end{equation}
In particular, it acts on the unit element and generators of $\mathfrak{g}$ as follows:
\begin{equation}
\Delta(1)=1 \otimes 1 \, , \quad\Delta(X_i)= X_i \otimes 1+1 \otimes X_i \qquad   (i=1,\dots,d) \, ,
\label{eq:gen}
\end{equation}
and extends to any monomial in $\mathcal{U}(\mathfrak{g})$ thanks to the homomorphism property in \eqref{eq:homom}, for example:
 \begin{align}
 \Delta(a_0 X_i X_j+b_0 X_k)&=a_0\Delta(X_{i})\Delta(X_{j})+b_0 \Delta(X_k) \nonumber \\
 &=a_0(X_i X_j \otimes 1+1 \otimes X_i X_j+X_i \otimes X_j+X_j \otimes X_i)+b_0 (X_k \otimes 1+1 \otimes X_k) \, .
 \label{eq:homo}
 \end{align}
The coproduct map preserves commutation relations $[X_i,X_j]=C_{ij}^k X_k$, in fact:
\begin{align}
[\Delta(X_i), \Delta(X_j)]=[X_i \otimes 1 + 1 \otimes X_i, X_j \otimes 1 + 1 \otimes X_j]&=[X_i,X_j] \otimes 1+ 1 \otimes [X_i,X_j] \nonumber\\
&=\Delta([X_i,X_j]) \nonumber \\
&=C_{ij}^k \Delta(X_k)\, ,
\label{eq:copmap}
\end{align}
it is therefore a homomorphism from $\mathcal{U}(\mathfrak{g})$ to $\mathcal{U}(\mathfrak{g}) \otimes \mathcal{U}(\mathfrak{g})$. Moreover, it is coassociative, i.e.:
\begin{equation}
 (\Delta \otimes id)  \circ \Delta=(id \otimes \Delta) \circ \Delta \, ,
\label{eq:cop}
\end{equation}
 as it can be proved straightforwardly acting on generators $X_i$. Given an element $U \in \mathcal{U}(\mathfrak{g})$, the map \eqref{eq:cop},  which is also a homomorphism from $\mathcal{U}(\mathfrak{g})$ to $\mathcal{U}(\mathfrak{g}) \otimes \mathcal{U}(\mathfrak{g}) \otimes \mathcal{U}(\mathfrak{g})$, allows to define elements in $\mathcal{U}(\mathfrak{g}) \otimes \mathcal{U}(\mathfrak{g}) \otimes \mathcal{U}(\mathfrak{g})$.
That said, the polynomial Casimir invariants in the sets \eqref{totalset} could be understood as the following elements in $\mathcal{U}(\mathfrak{g}) \otimes \mathcal{U}(\mathfrak{g}) \otimes \mathcal{U}(\mathfrak{g})$ $($using Sweedler's notation$)$:
\begin{align}
C_1^{(r)} &= C^{(r)} \otimes 1 \otimes 1\, , \qquad\,\, C_2^{(r)} =  1 \otimes  C^{(r)} \otimes 1 \, , \qquad\,\, C_3^{(r)} =  1 \otimes 1 \otimes  C^{(r)}  \\
C^{(r)}_{12} &= C_{(1)}^{(r)} \otimes C_{(2)}^{(r)}  \otimes 1 \, , \quad C^{(r)}_{23} = 1 \otimes C_{(1)}^{(r)}  \otimes C_{(2)}^{(r)}  \, , \quad C^{(r)}_{13} =  C_{(1)}^{(r)}  \otimes 1 \otimes C_{(2)}^{(r)}
\end{align}
and $C_{123}^{(r)}=C_{(1)}^{(r)}  \otimes C_{(2)}^{(r)} \otimes C_{(3)}^{(r)}$, with $r=1, \dots, N_p(\mathfrak{g})$. From this perspective, in order to implement computations at the level of the computer algebra software\footnote{We use \textsf{Mathematica}\textsuperscript{\tiny{\textregistered}}, specifically the package \textsf{NCAlgebra}.}, one can rely on the following identifications:
\begin{align}
X_i^{[\alpha]} \quad &\longleftrightarrow \quad 1^{\otimes (\alpha-1)} \otimes X_i \otimes  1^{\otimes (3-\alpha)}\\
X_i^{[\alpha, \beta]}\quad &\longleftrightarrow \quad 1^{\otimes (\alpha-1)} \otimes X_i \otimes  1^{\otimes (3-\alpha)}+ 1^{\otimes (\beta-1)} \otimes X_i \otimes  1^{\otimes (3-\beta)}\\
X_i^{[1, 2, 3]} \quad &\longleftrightarrow\quad X_i\otimes 1 \otimes 1 + 1 \otimes X_i \otimes 1 + 1 \otimes 1 \otimes X_i \, .
\label{eq:ident}
\end{align}
\label{rem}
\end{remark}
\noindent These are the Casimir elements that we will be considering throughout the paper to construct polynomial algebras. In particular, to define closure we now introduce the following right-nested commutators:
\begin{align}
C_{\alpha_1 \alpha_2 \alpha_3 \alpha_4}^{(r,s)}&:=[C_{\alpha_1 \alpha_2}^{(r)},C_{\alpha_3 \alpha_4}^{(s)}] \hskip 5.85cm r,s =1, \dots, N_p(\mathfrak{g}) \\
C_{\alpha_1 \alpha_2 \alpha_3 \alpha_4 \alpha_5 \alpha_6}^{(r,s,t)}&:=[C_{\alpha_1 \alpha_2}^{(r)}, C_{\alpha_3 \alpha_4 \alpha_5 \alpha_6}^{(s,t)}]=[C_{\alpha_1 \alpha_2}^{(r)},[C_{\alpha_3\alpha_4}^{(s)},C_{\alpha_5 \alpha_6}^{(t)}]] \hskip 1cm r,s,t=1, \dots, N_p(\mathfrak{g}) \, ,
\end{align}
\noindent and more in general:
\begin{equation}
C_{\alpha_1 \alpha_2 \dots \alpha_{2n}}^{(r_1,\dots,r_{n})}:=[C^{(r_1)}_{\alpha_1 \alpha_2},C_{\alpha_3 \alpha_4 \dots \alpha_{2n}}^{(r_2,  \dots,r_{n} )} ]= [ C^{(r_1)}_{\alpha_1 \alpha_2},[C^{(r_2)}_{\alpha_3 \alpha_4},\dots [C^{(r_{n-1})}_{\alpha_{2n-3}\alpha_{2n-2} },C^{(r_n)}_{\alpha_{2n-1}\alpha_{2n}}]\dots]] \, .
\label{gen}
\end{equation}
Each pair of indices takes values in the set $\{(12),(23),(13)\}$, i.e.  $(\alpha_{2k-1} \alpha_{2k})  \in \{(12),(23),(13)\}$, $k=1, \dots, n$.
\noindent We then define a commutator algebra in terms of the algebra relations among the nested commutators for a given $n \geq 2$ and all lower degree nested commutators. In particular, if closure is obtained in terms of the elements:
\begin{equation}
\mathcal{C}^{(r)} \cup  \{C^{(r,s)}_{\alpha_1 \alpha_2 \alpha_3 \alpha_4}\} \qquad r,s =1, \dots, N_p(\mathfrak{g}) \, ,
\label{eq:closure1}
\end{equation}
\noindent  we will be dealing with polynomial algebras of the form:
 \begin{equation}
[C_{\alpha _1\alpha_2}^{(r)},[C_{\alpha_3 \alpha_4}^{(s)},C_{\alpha_5 \alpha_6}^{(t)}]]=P_{\alpha_1 \dots \alpha_6 }(C^{(u)}_{\alpha_7 \alpha_8}) \, ,
\label{exp}
\end{equation}

\noindent where $P_{\alpha_1 \dots \alpha_6 }$ are polynomials of some degree in the generators \eqref{totalset} and $r,s,t,u=1, \dots, N_p(\mathfrak{g})$.
\begin{remark}
In many examples \eqref{exp} will be just re-expressed as a quadratic expansion in the Casimir elements belonging to the set \eqref{totalset}. In this case, we say that the polyomial algebra closes quadratically.
In general, the structure constants may depend on the one-index Casimir invariants $C_\alpha^{(r)}$, the total Casimir invariants $C^{(r)}_{123}$ and other central elements that may arise for the specific Lie algebra we are considering.
\end{remark}
\noindent If closure is not achieved at this level, we then consider higher order right-nested commutators:
\begin{equation}
\mathcal{C}^{(r_1)} \cup  \{C^{(r_1,r_2)}_{\alpha_1 \alpha_2  \alpha_3 \alpha_4}\}  \cup \dots \cup  \{C^{(r_1, \dots, r_n)}_{\alpha_1 \alpha_2  \dots  \alpha_{2n-1}\alpha_{2n}}\}  \, ,
\label{eq:closure2}
\end{equation}
for some $n > 2$ until closure is achieved in terms of all lower degree ones. The finitely generated polynomial algebras are then defined as the finitely generated commutator algebras obtained from the Casimir invariants in the set \eqref{totalset} and the additional right-nested commutators \eqref{gen} for some $n=n^*$ necessary in order to get closure.

All computations are performed abstractly, with the help of the  \textsf{Mathematica}\textsuperscript{\tiny{\textregistered}} package \textsf{NCAlgebra}, by relying to the underlying Lie algebra generators $X_i^{[\alpha]} \in \mathfrak{g}^{[\alpha]}$ and their abstract commutation relations. More specifically, we will be working on the universal enveloping algebra $\mathcal{U}(\mathfrak{g} \oplus \mathfrak{g} \oplus \mathfrak{g})$.
Thus, taken three copies of the same Lie algebra $\mathfrak{g}$ of dimension $d:=\text{dim}(\mathfrak{g})$ spanned by the basis generators $X_i^{[\alpha]}$, we will be dealing with ordered elements of the type:
\begin{equation}
\bigl(X_1^{[1]}\bigl)^{a_1} \dots \bigl(X_d^{[1]}\bigl)^{a_d} \bigl(X_1^{[2]}\bigl)^{b_1} \dots \bigl(X_d^{[2]}\bigl)^{b_d} \bigl(X_1^{[3]}\bigl)^{c_1} \dots \bigl(X_d^{[3]}\bigl)^{c_d}  \qquad a_i,b_i,c_i \in  \mathbb{Z}_{\geq 0}\, .
\label{basis}
\end{equation}

\noindent In what follows, we present a series of explicit examples of algebraic structures obtained when we apply the construction to different types of Lie algebras, such as \emph{simple}, \emph{nilpotent}, \emph{solvable} or \emph{Levi decomposable} ones. All the examples discussed in this paper come from - and follow the notations of - the classification reported in \cite{sno14}, where lists of all indecomposable Lie algebras up to dimension six are given. In particular, following this reference, we will introduce each Lie algebra $\mathfrak{g}$ in a given basis together with its associated commutation table and Casimir invariants.

\section{Simple Lie algebras}
\label{sec3}
Just to fix the notations, and lay the basis to understand how the construction actually works, we will begin by briefly discussing some examples of \emph{simple Lie algebras}, i.e. those Lie algebras that do not possess any nontrivial ideal\footnote{We recall that an ideal $\mathfrak{i}$ is a subalgebra of $\mathfrak{g}$ such that $[\mathfrak{i}, \mathfrak{g}] \subseteq \mathfrak{i}$. The Lie algebra $\mathfrak{g}$ itself and $\{0\}$ are trivial ideals \cite{sno14}.}, starting from the $\mathfrak{sl}_2(\mathbb{F})$ case. 
\subsection{From the simple Lie algebra $\boldsymbol{\mathfrak{sl}_{2}}(\mathbb{F})$ to the Racah algebra $R(3)$}
\label{sec3.1}

\noindent As a first illustrative example let us consider again the Lie algebra $\mathfrak{sl}_2(\mathbb{F})$ in the basis generators $X_i \equiv \{X_1, X_2, X_3\}$ with commutation table \cite{sno14}:

\begin{center}
	\begin{tabular}{| l | c|c | r| }
		\hline
		& $X_1$ & $X_2$ & $X_3$ \\ \hline
		$X_1$ & 0 & 2$X_1$ &$-X_2$ \\ \hline
		$X_2$ & $-2X_1$ & 0 & $2X_3$\\ \hline
		$X_3$ & $X_2$ & $-2X_3$ & 0\\
		\hline
	\end{tabular}
\end{center}

\noindent As previously showed, this algebra is endowed with the Casimir element ($r=1$):  
\begin{equation}
C^{(1)}(X_1,X_2,X_3)= X_2^2+4 X_1X_3+2X_2 \, .
\end{equation}

\noindent Now, let us consider three copies of the same simple Lie algebra $\mathfrak{sl}_2(\mathbb{F})$, each copy characterized by its own basis generators $\{X_i^{[\alpha]}\}$ ($\alpha=1,2,3$). By following the general procedure we described in the previous section let us introduce the elements:
\begin{align}
&C^{(1)}_\alpha=\bigl(X_2^{[\alpha]}\bigl)^2+4 X_1^{[\alpha]} X_3^{[\alpha]}+2 X_2^{[\alpha]}\\
&C_{\alpha \beta}^{(1)}=\bigl(X_2^{[\alpha,\beta]}\bigl)^2+4 X_1^{[\alpha,\beta]} X_3^{[\alpha,\beta]}+2 X_2^{[\alpha, \beta]} \, ,
\end{align}
together with: 
\begin{equation}
C_{123}^{(1)}=\bigl(X_2^{[1,2,3]}\bigl)^2+4 X_1^{[1,2,3]} X_3^{[1,2,3]}+2 X_2^{[1,2,3]} \, .
\end{equation}
These elements are linearly dependent, as the following linear relation holds:
\begin{equation}
C^{(1)}_{123}=C^{(1)}_{12}+C^{(1)}_{23}+C^{(1)}_{13}-C^{(1)}_1-C^{(1)}_2-C^{(1)}_3 \, .
\label{eq:from}
\end{equation}
This relation, as the other ones we will be obtaining in the paper, has to be understood as an equality arising in terms of monomials involving the generators $X_i^{[\alpha]}$. This means that the quadratic combination:
\begin{align}
\bigl(X_2^{[1,2,3]}\bigl)^2+4 X_1^{[1,2,3]} X_3^{[1,2,3]}+2 X_2^{[1,2,3]}
\end{align}
namely the left hand side of \eqref{eq:from}, once expanded is equivalent to the following one:
\begin{align}
&\bigl(X_2^{[1,2]}\bigl)^2+4 X_1^{[1,2]} X_3^{[1,2]}+2 X_2^{[1,2]} +\bigl(X_2^{[2,3]}\bigl)^2+4 X_1^{[2,3]} X_3^{[2,3]}+2 X_2^{[2,3]}+\bigl(X_2^{[1,3]}\bigl)^2+4 X_1^{[1,3]} X_3^{[1,3]}+2 X_2^{[1,3]}
\label{eq:ablingen}\nonumber\\
&-\bigl(\bigl(X_2^{[1]}\bigl)^2+4 X_1^{[1]} X_3^{[1]}+2 X_2^{[1]}\bigl)-\bigl(\bigl(X_2^{[2]}\bigl)^2+4 X_1^{[2]} X_3^{[2]}+2 X_2^{[2]}\bigl)-\bigl(\bigl(X_2^{[3]}\bigl)^2+4 X_1^{[3]} X_3^{[3]}+2 X_2^{[3]}\bigl) \, ,
\end{align}
i.e. the right hand side of \eqref{eq:from}. This also extends to abstract commutation relations, that will be obtained by reconstructing terms from the initial Lie algebras generators and their abstract commutation relations \eqref{eq11}-\eqref{eq12}. We thus have a set of Casimir invariants: 
\begin{equation}
\mathcal{C}^{(1)}:=\{C^{(1)}_1,C^{(1)}_2,C^{(1)}_3,C^{(1)}_{12},C^{(1)}_{13},C^{(1)}_{23},C^{(1)}_{123}\}\, .
\label{set}
\end{equation}
The only non-zero commutation relations in the set $\mathcal{C}^{(1)}$ are those arising from the two indices generators $C^{(1)}_{\alpha \beta}$, and clearly they cannot be expressed back in terms of the original elements in $\mathcal{C}^{(1)}$.  Instead, they can be used to define the new elements:
\begin{align}
C^{(1,1)}_{1223}&:= [C^{(1)}_{12}, C^{(1)}_{23}]=-C^{(1,1)}_{2312} \label{1a}\\
C^{(1,1)}_{2313}&:=[C^{(1)}_{23},C^{(1)}_{13}]=-C^{(1,1)}_{1323}\label{1b}\\
C^{(1,1)}_{1312}&:=[C^{(1)}_{13},C^{(1)}_{12}]=-C^{(1,1)}_{1213} \, .
\label{1c}
\end{align}

\noindent At this point, we check if these three elements are independent. To do so, we impose:
\begin{equation}
a_1 C_{1223}^{(1,1)}+a_2 C_{2313}^{(1,1)}+a_3 C_{1312}^{(1,1)}=0 \, .
\label{ind}
\end{equation}
By expanding this equation in the original generators $X_i^{[\alpha]}$ and recollecting terms, we obtain an overdetermined system composed by six equations for the parameters $a_i$. Each equation in the system is given in terms of the unique constraint $a_1+a_2+a_3=0$, that is solved for $a_3=-a_1-a_2$. In this way we get the relation:
\begin{equation}
a_1 C_{1223}^{(1,1)}+a_2 C_{2313}^{(1,1)}-(a_1+a_2) C_{1312}^{(1,1)}=0 \, .
\label{indep}
\end{equation}
This equation gives us the two equalities:
\begin{equation}
C_{1223}^{(1,1)}=C_{1312}^{(1,1)} \qquad C_{2313}^{(1,1)}=C_{1312}^{(1,1)} \, ,
\label{eq:equalities}
\end{equation}
which combined together lead to:
\begin{equation} C_{1223}^{(1,1)}=C_{2313}^{(1,1)}=C_{1312}^{(1,1)} \, .
\label{equal}
\end{equation}
\begin{remark}
 We notice that, because of the relation \eqref{eq:from}, the chain of equalities \eqref{equal} directly arise by taking the commutator of $C^{(1)}_{123}$ with $C^{(1)}_{\alpha \beta}$:
\begin{align}
&0=[C^{(1)}_{123},C^{(1)}_{12}]=[C^{(1)}_{13}+C^{(1)}_{23}, C^{(1)}_{12}]=[C^{(1)}_{13}, C^{(1)}_{12}]+[C^{(1)}_{23}, C^{(1)}_{12}]=C^{(1,1)}_{1312}+C^{(1,1)} _{2312}\\
&0=[C^{(1)}_{123},C^{(1)}_{13}]=[C^{(1)}_{12}+C^{(1)}_{23}, C^{(1)}_{13}]=[C^{(1)}_{12}, C^{(1)}_{13}]+[C^{(1)}_{23}, C^{(1)}_{13}]=C^{(1,1)}_{1213}+C^{(1,1)}_{2313}\\
&0=[C^{(1)}_{123},C^{(1)}_{23}]=[C^{(1)}_{12}+C^{(1)}_{13}, C^{(1)}_{23}]=[C^{(1)}_{12}, C^{(1)}_{23}]+[C^{(1)}_{13}, C^{(1)}_{23}] =C^{(1,1)}_{1223}+C^{(1,1)}_{1323}
\end{align}
which, taking into account \eqref{1a}-\eqref{1c}, result in \eqref{equal}.
\end{remark}

\noindent Thus, an additional element can be introduced as:
\begin{equation}
C_{1223}^{(1,1)}= [C^{(1)}_{12}, C^{(1)}_{23}]= [C^{(1)}_{23}, C^{(1)}_{13}]=[C^{(1)}_{13}, C^{(1)}_{12}]\, .
\label{eq:add}
\end{equation}
 This element, together with the ones in the set \eqref{set},  close in the quadratic algebra:
\begin{align}
&[C_{12}^{(1)}, C_{1223}^{(1,1)}]=8\bigl(C^{(1)}_{23}C^{(1)}_{12}-C^{(1)}_{12}C^{(1)}_{13}+(C^{(1)}_2-C^{(1)}_1)(C^{(1)}_3-C^{(1)}_{123})\bigl) \label{1r}\\
&[C_{23}^{(1)}, C_{1223}^{(1,1)}]=8\bigl(C^{(1)}_{13}C^{(1)}_{23}-C^{(1)}_{23}C^{(1)}_{12}+(C^{(1)}_3-C^{(1)}_2)(C^{(1)}_1-C^{(1)}_{123})\bigl)\label{2r}\\
&[C_{13}^{(1)}, C_{1223}^{(1,1)}]=8\bigl(C^{(1)}_{12}C^{(1)}_{13}-C^{(1)}_{13}C^{(1)}_{23}+(C^{(1)}_1-C^{(1)}_3)(C^{(1)}_2-C^{(1)}_{123})\bigl) \label{3r}\, .
\end{align}
 Thus, in this case, the nested commutators $[C^{(1)}_{\alpha_1 \alpha_2}, [C^{(1)}_{\alpha_3 \alpha_4}, C^{(1)}_{\alpha_5 \alpha_6}]]$ turn out to be expressible as quadratic combinations of the Casimir invariants appearing in the set \eqref{set}.  Moreover, the following relation holds:
 \begin{equation}
 [C_{12}^{(1)},C_{1223}^{(1,1)}]+  [C_{23}^{(1)}, C_{1223}^{(1,1)}]+  [C_{13}^{(1)},  C_{1223}^{(1,1)}] = 0\, .
 \label{rel}
 \end{equation}
 
 \noindent  This result was expected as this quadratic algebra is in fact related to the rank-$1$ Racah algebra $R(3)$, $C^{(1)}_{1}$, $C^{(1)}_2$, $C^{(1)}_3$ and $C^{(1)}_{123}$ playing the role of central elements \cite{bie2020racah, https://doi.org/10.48550/arxiv.2105.01086}. 

\begin{remark}
The associated algebraic Hamiltonian of the finitely generated Racah algebra is given by the total Casimir invariant $C^{(1)}_{123}$ and the set $\{C^{(1)}_1,C^{(1)}_2,C^{(1)}_3,C^{(1)}_{12},C^{(1)}_{13},C^{(1)}_{23},C^{(1,1)}_{1223}\}$ can be also interpreted as its commutant. 
	\label{eq:re}
\end{remark}

\subsection{The simple Lie algebra $\mathfrak{so}(1,3)$}
 \label{sec3.2}

Let us now focus on another simple Lie algebra, namely the six-dimensional Lie algebra $\mathfrak{so}(1,3)$ in the basis generators $X_i \equiv \{X_1,X_2,X_3,X_4,X_5,X_6\}$ with  commutation table:
\begin{center}
	\begin{tabular}{| l | c|c|c|c | c|r| }
		\hline
		& $X_1$ & $X_2$ & $X_3$ &$X_4$&$X_5$ &$X_6$\\ \hline
		$X_1$ & 0 & $X_3$ &$-X_2$ &0&$X_6$&$-X_5$\\ \hline
		$X_2$ & $-X_3$& 0 & $X_1$&$-X_6$&0&$X_4$\\ \hline
		$X_3$ & $X_2$ & $-X_1$ & 0&$X_5$&$-X_4$&$0$\\ \hline
		$X_4$ & $0$ & $X_6$ & $-X_5$&$0$&$-X_3$&$X_2$\\ \hline
		$X_5$ & $-X_6$ & $0$ & $X_4$&$X_3$&0&$-X_1$\\ \hline
		$X_6$ & $X_5$ & $-X_4$ & $0$&$-X_2$&$X_1$&$0$ \\\hline
	\end{tabular}
\end{center}
\noindent In this case $N(\mathfrak{g})=N_p(\mathfrak{g})=6-4=2$ and to obtain the quadratic Casimir invariants we need to find the two independent solutions $F_{1,2}(x_1,x_2,x_3,x_4,x_5,x_6)$ of the following system of PDEs:
\begin{equation}
\begin{cases}
\hat{X_1}F&=x_3 F_{x_2}-x_2 F_{x_3}+x_6 F_{x_5}-x_5 F_{x_6} =0
\\
\hat{X_2}F&=-x_3 F_{x_1}+x_1 F_{x_3}-x_6 F_{x_4}+x_4 F_{x_6} =0
\\
\hat{X_3}F&=x_2 F_{x_1}-x_1 F_{x_2}+x_5 F_{x_4}-x_4 F_{x_5} =0 
\\
\hat{X_4}F&=x_6 F_{x_2}-x_5 F_{x_3}-x_3 F_{x_5}+x_2 F_{x_6} =0 
\\
\hat{X_5}F&=-x_6 F_{x_1}+x_4 F_{x_3}+x_3 F_{x_4}-x_1 F_{x_6} =0 
\\
\hat{X_6}F&=x_5 F_{x_1}-x_4 F_{x_2}-x_2F_{x_4}+x_1 F_{x_5}=0 \, . 
\end{cases}
\end{equation} 

\noindent These two polynomial solutions can be taken as \cite{sno14}:
\begin{equation}
\begin{cases}
F_1(x_1,x_2,x_3,x_4,x_5,x_6)=x_1 x_4+x_2 x_5+x_3 x_6 \\
F_2(x_1,x_2,x_3,x_4,x_5,x_6) = x_1^2+ x_2^2+x_3^2 -x_4^2-x_5^2- x_6^2 \, .
\end{cases}
\label{[q:sol}
\end{equation}
From them, we can obtain the following quadratic Casimir invariants in $\mathcal{U}(\mathfrak{so}(1,3))$:
\begin{equation}
\small C^{(1)}(X_1, \dots, X_6) = X_1 X_4+X_2 X_5+X_3 X_6 \, , \quad C^{(2)} (X_1, \dots, X_6) = X_1^2+X_2^2+X_3^2-X_4^2-X_5^2-X_6^2 \, .
\label{eq:casimir}
\end{equation}
\noindent Let us consider  three copies of the same Lie algebra with generators $\{X_i^{[\alpha]}\}$, $i=1,2,3,4,5,6$ and $\alpha=1,2,3$ respectively. From the above elements, we construct the following Casimir invariants:
\begin{align}
\mathcal{C}^{(1)}&:=\{C^{(1)}_1, C^{(1)}_2,C^{(1)}_3,C^{(1)}_{12},C^{(1)}_{23},C^{(1)}_{13},C^{(1)}_{123}\}  \label{eq1} \\
\mathcal{C}^{(2)}&:=\{C^{(2)}_1, C^{(2)}_2,C^{(2)}_3,C^{(2)}_{12},C^{(2)}_{23},C^{(2)}_{13},C^{(2)}_{123}\} \, ,
\label{eq2}
\end{align} 
where the one and two indices elements are given by:
\begin{align}
&C^{(1)}_\alpha=X_1^{[\alpha]}X_4^{[\alpha]}+X_2^{[\alpha]}X_5^{[\alpha]}+X_3^{[\alpha]}X_6^{[\alpha]}\\
&C^{(2)}_\alpha=\bigl(X_1^{[\alpha]}\bigl)^2+\bigl(X_2^{[\alpha]}\bigl)^2+\bigl(X_3^{[\alpha]}\bigl)^2-\bigl(X_4^{[\alpha]}\bigl)^2-\bigl(X_5^{[\alpha]}\bigl)^2-\bigl(X_6^{[\alpha]}\bigl)^2\\
&C_{\alpha \beta}^{(1)}=X_1^{[\alpha,\beta]}X_4^{[\alpha,\beta]}+X_2^{[\alpha,\beta]}X_5^{[\alpha,\beta]}+X_3^{[\alpha,\beta]}X_6^{[\alpha,\beta]}\\
&C^{(2)}_{\alpha \beta}=\bigl(X_1^{[\alpha,\beta]}\bigl)^2+\bigl(X_2^{[\alpha,\beta]}\bigl)^2+\bigl(X_3^{[\alpha,\beta]}\bigl)^2-\bigl(X_4^{[\alpha,\beta]}\bigl)^2-\bigl(X_5^{[\alpha,\beta]}\bigl)^2-\bigl(X_6^{[\alpha,\beta]}\bigl)^2 \, ,
\end{align} 
whereas the three indices ones read:
\begin{align}
C^{(1)}_{123}&=X_1^{[1,2,3]}X_4^{[1,2,3]}+X_2^{[1,2,3]}X_5^{[1,2,3]}+X_3^{[1,2,3]}X_6^{[1,2,3]}\\
C^{(2)}_{123}&=\bigl(X_1^{[1,2,3]}\bigl)^2+\bigl(X_2^{[1,2,3]}\bigl)^2+\bigl(X_3^{[1,2,3]}\bigl)^2-\bigl(X_4^{[1,2,3]}\bigl)^2-\bigl(X_5^{[1,2,3]}\bigl)^2-\bigl(X_6^{[1,2,3]}\bigl)^2 \, .
\label{eq:casfull}
\end{align}
Again, a direct computation shows that the following linear relations hold for both sets:
\begin{align}
C^{(r)}_{123}&=C^{(r)}_{12}+C^{(r)}_{13}+C^{(r)}_{23}-C^{(r)}_1-C^{(r)}_2-C^{(r)}_3 \, \qquad (r=1,2) \, . 
\label{linrelr}
\end{align}
Moreover, the total Casimir invariants $C^{(r)}_{123}$ and the one index ones $C_{\alpha}^{(r)}$ commute each other and with all the other two indices Casimir invariants in the above sets. Therefore,
in this case, the following additional elements can be constructed:
\begin{align}
C^{(1,1)}_{1223}&:= [C^{(1)}_{12}, C^{(1)}_{23}]=-C^{(1,1)}_{2312}
\qquad C^{(2,2)}_{1223}:= [C^{(2)}_{12}, C^{(2)}_{23}]=-C^{(2,2)}_{2312}\\
C^{(1,1)}_{1312}&:=[C^{(1)}_{13},C^{(1)}_{12}]=-C^{(1,1)}_{1213}  \qquad
C^{(2,2)}_{1312}:=[C^{(2)}_{13},C^{(2)}_{12}]=-C^{(2,2)}_{1213}\\
C^{(1,1)}_{2313}&:=[C^{(1)}_{23},C^{(1)}_{13}]=-C^{(1,1)}_{1323} \qquad
C^{(2,2)}_{2313}:=[C^{(2)}_{23},C^{(2)}_{13}]=-C^{(2,2)}_{1323} \, ,
\label{eq:aps}
\end{align}
together with:
\begin{align}
C^{(1,2)}_{1223}&:= [C^{(1)}_{12}, C^{(2)}_{23}]=-C^{(2,1)}_{2312}  \qquad C^{(2,1)}_{1223}:= [ C^{(2)}_{12}, C^{(1)}_{23}]=-C^{(1,2)}_{2312} \\
C^{(1,2)}_{2313}&:=[C^{(1)}_{23},C^{(2)}_{13}]=-C^{(2,1)}_{1323}\qquad C^{(2,1)}_{2313}:= [ C^{(2)}_{23}, C^{(1)}_{13}]=-C^{(1,2)}_{1323}\\
C^{(1,2)}_{1312}&:=[C^{(1)}_{13},C^{(2)}_{12}]=-C^{(2,1)}_{1213} \qquad C^{(2,1)}_{1312}:= [ C^{(2)}_{13}, C^{(1)}_{12}]=-C^{(1,2)}_{1213}  \, ,
\end{align}
being [$C^{(1)}_{12}, C^{(2)}_{12}]=[C^{(1)}_{13},C^{(2)}_{13}]=[C^{(1)}_{23},C^{(2)}_{23}]=0$.  Now, by imposing:
\begin{align}
&a_1 C^{(1,1)}_{1223}+a_2 C^{(1,1)}_{2313}+a_3 C^{(1,1)}_{1312}+a_4 C^{(2,2)}_{1223}+a_5 C^{(2,2)}_{2313}+a_6 C^{(2,2)}_{1312}+ \nonumber \\
&a_7 C^{(1,2)}_{1223}+a_8 C^{(1,2)}_{2313}+a_9 C^{(1,2)}_{1312}+a_{10} C^{(2,1)}_{1223}+a_{11} C^{(2,1)}_{2313}+a_{12} C^{(2,1)}_{1312}=0 \, ,
\label{expa}
\end{align}
expanding the expression in terms of the original generators $X_i^{[\alpha]}$ and recollecting equal terms, we obtain an overdetermined system composed by forty-eight equations for the coefficients $a_i$. All the equations are solved for: 
\begin{equation}
a_6=1/4 (a_1 + a_2 + a_3) -  a_4 - a_5 \, , \qquad a_{12} = -a_{11} - a_{10} - a_9 - a_8 - a_7 \, .
\label{eq:constraints}
\end{equation}
We thus get the following equalities:
\begin{align}
C^{(1,1)}_{1223}&=C^{(1,1)}_{2313}=C^{(1,1)}_{1312} =-C^{(2,2)}_{1223}/4=-C^{(2,2)}_{2313}/4=-C^{(2,2)}_{1312}/4 \\ 
C^{(1,2)}_{1223}&=C_{2313}^{(1,2)}=C^{(1,2)}_{1312}=C_{1223}^{(2,1)}=C_{2313}^{(2,1)}=C^{(2,1)}_{1312}\, .
\label{eq:interrelations}
\end{align}
Because of these relations, we are led to introduce the two independent elements:
{\small\begin{align}
C_{1223}^{(1,1)}&:=[C^{(1)}_{12}, C^{(1)}_{23}]=[C^{(1)}_{23}, C^{(1)}_{13}]=[C^{(1)}_{13}, C^{(1)}_{12}]=[C^{(2)}_{23}, C^{(2)}_{12}]/4=[C^{(2)}_{13}, C^{(2)}_{23}]/4=[C^{(2)}_{13}, C^{(2)}_{12}]/4 \label{el1}\\
C_{1223}^{(1,2)}&:=[C^{(1)}_{12}, C^{(2)}_{23}]=[C^{(1)}_{23}, C^{(2)}_{13}]=[C^{(1)}_{13}, C^{(2)}_{12}]=[C^{(2)}_{12}, C^{(1)}_{23}]=[C^{(2)}_{23}, C^{(1)}_{13}]=[C^{(2)}_{13}, C^{(1)}_{12}] \, .
\label{el2}
\end{align}}

\noindent The above elements, together with the Casimir invariants in the two sets \eqref{eq1}-\eqref{eq2}, will be the defining generators of the polynomial algebra.   First of all, let us notice that they  commute each other, i.e.:
\begin{equation}
[C^{(1,1)}_{1223}, C_{1223}^{(1,2)}]=0 \, .
\label{eq:zerocomm}
\end{equation}
Moreover, together with the other Casimir invariants, they close in the following quadratic algebra:
\begin{align}
	 [C_{12}^{(1)},C^{(1,1)}_{1223}]&=1/2\bigl( C^{(1)}_{12}(C^{(2)}_{23}-C^{(2)}_{13})+(C^{(1)}_{23}-C^{(1)}_{13})C^{(2)}_{12}\bigl)\nonumber\\
	&\hskip 0.25cm+1/2\bigl((C^{(1)}_1-C^{(1)}_2)(C^{(2)}_{123}-C^{(2)}_3)+(C^{(1)}_{123}-C^{(1)}_3)(C^{(2)}_1-C^{(2)}_2)\bigl)\\
	 [C_{23}^{(1)},C^{(1,1)}_{1223}]&=1/2\bigl(C^{(1)}_{23}(C^{(2)}_{13}-C^{(2)}_{12})+(C^{(1)}_{13}-C^{(1)}_{12})C^{(2)}_{23}\bigl) \nonumber \\
	 &\hskip 0.25cm+1/2\bigl((C^{(1)}_2-C^{(1)}_3)(C^{(2)}_{123}-C^{(2)}_1)+(C^{(1)}_{123}-C^{(1)}_1)(C^{(2)}_2-C^{(2)}_3)\bigl)\\
	[C_{13}^{(1)},C^{(1,1)}_{1223}]&=1/2\bigl(C^{(1)}_{13}(C^{(2)}_{12}-C^{(2)}_{23})+(C^{(1)}_{12}-C^{(1)}_{23})C^{(2)}_{13}\bigl)\nonumber\\
	&\hskip 0.25cm-1/2\bigl((C^{(1)}_1-C^{(1)}_3)(C^{(2)}_{123}-C^{(2)}_2)-(C^{(1)}_{123}-C^{(1)}_2)(C^{(2)}_1-C^{(2)}_3)\bigl) \, ,
	\end{align}
	\begin{align}
	[C_{12}^{(2)},C^{(1,1)}_{1223}]&=2 \bigl(C^{(1)}_{12}C^{(1)}_{13}-C^{(1)}_{23}C^{(1)}_{12}+(C^{(1)}_1-C^{(1)}_2)(C^{(1)}_3-C^{(1)}_{123})\bigl)\nonumber\\
	&\hskip 0.25cm-1/2\bigl(C^{(2)}_{12}C^{(2)}_{13}-C^{(2)}_{23}C^{(2)}_{12}+(C^{(2)}_1-C^{(2)}_2)(C^{(2)}_3-C^{(2)}_{123})\bigl)\\
	[C_{23}^{(2)},C^{(1,1)}_{1223}]&=2 \bigl(C^{(1)}_{23} C^{(1)}_{12}-C^{(1)}_{13}C^{(1)}_{23}+(C^{(1)}_2-C^{(1)}_3)(C^{(1)}_1-C^{(1)}_{123})\bigl)\nonumber\\
	&\hskip 0.25cm-1/2\bigl(C^{(2)}_{23} C^{(2)}_{12}-C^{(2)}_{13}C^{(2)}_{23}+(C^{(2)}_2-C^{(2)}_3)(C^{(2)}_1-C^{(2)}_{123})\bigl)\\
	[C_{13}^{(2)},C^{(1,1)}_{1223}]&=2 \bigl(C^{(1)}_{13} C^{(1)}_{23}-C^{(1)}_{12}C^{(1)}_{13}-(C^{(1)}_1-C^{(1)}_3)(C^{(1)}_2-C^{(1)}_{123})\bigl)\nonumber\\
	&\hskip 0.25cm-1/2\bigl(C^{(2)}_{13} C^{(2)}_{23}-C^{(2)}_{12}C^{(2)}_{13}-(C^{(2)}_1-C^{(2)}_3)(C^{(2)}_2-C^{(2)}_{123})\bigl) \, ,
\label{key}	\end{align}
\begin{align}
  [C_{12}^{(1)},C^{(1,2)}_{1223}]&=[C_{12}^{(2)},C^{(1,1)}_{1223}] \qquad [C_{12}^{(2)},C^{(1,2)}_{1223}] =-4[C_{12}^{(1)},C^{(1,1)}_{1223}]\\ [C_{23}^{(1)},C^{(1,2)}_{1223}]&=[C_{23}^{(2)},C^{(1,1)}_{1223}]\qquad [C_{23}^{(2)},C^{(1,2)}_{1223}] =-4[C_{23}^{(1)},C^{(1,1)}_{1223}]\\ [C_{13}^{(1)},C^{(1,2)}_{1223}]&=[C_{13}^{(2)},C^{(1,1)}_{1223}] \qquad [C_{13}^{(2)},C^{(1,2)}_{1223}]=-4[C_{13}^{(1)},C^{(1,1)}_{1223}]\, ,
\end{align}
with:
\begin{equation}
[C_{12}^{(1)},C^{(1,1)}_{1223}]+[C_{23}^{(1)},C^{(1,1)}_{1223}]+[C_{13}^{(1)},C^{(1,1)}_{1223}]=0 \, , \quad  [C_{12}^{(1)},C^{(1,2)}_{1223}]+[C_{23}^{(1)},C^{(1,2)}_{1223}]+[C_{13}^{(1)},C^{(1,2)}_{1223}]=0 \, .
\label{addrel}
\end{equation}
\noindent At least formally, this quadratic algebra seems to share some common features with the one constructed from the simple Lie algebra $\mathfrak{sl}_2$. Again, quadratic Casimir invariants belonging to a given set turn out to be linearly dependent and to satisfy the same linear relation. Also, all nested commutators $[C^{(1)}_{\alpha_1 \alpha_2}, [C^{(1)}_{\alpha_3  \alpha_4}, C^{(1)}_{\alpha_5  \alpha_6}]]$ turn out to be expressible as quadratic combinations of the Casimir invariants, however in this case appearing in both the sets \eqref{eq1}-\eqref{eq2}. This holds true also for the additional nested commutators $[C^{(1)}_{\alpha_1  \alpha_2}, [C^{(1)}_{\alpha_3  \alpha_4}, C^{(2)}_{\alpha_5 \alpha_6}]]$. 

The presence of two Casimir elements for the initial simple Lie algebra makes the quadratic structure more involved, as mixed terms involving intermediate Casimir invariants of both sets appear.  
\section{Nilpotent Lie algebras}
\label{sec4}

This Section \ref{sec4} is devoted to the analysis of some representatives of \emph{nilpotent Lie algebras}. 	We recall that a Lie algebra $\mathfrak{g}$ is called \emph{nilpotent} if the lower central series terminates \cite{sno14}, i.e. if there exists a $k \in  \mathbb{N}$ such that $\mathfrak{g}^k = 0$, where the lower central series $\mathfrak{g}=\mathfrak{g}^1 \supseteq \dots\supseteq  \mathfrak{g}^k \supseteq \dots $ is defined recursively: $\mathfrak{g}^k=[\mathfrak{g}^{k-1},\mathfrak{g}]$, $\mathfrak{g}^1=\mathfrak{g}$.
 \subsection{ The nilpotent Lie algebra $\boldsymbol{\mathfrak{n}_{5,5}}$}
  \label{sec4.1}

\noindent  Let us consider the five-dimensional nilpotent Lie algebra $\mathfrak{n}_{5,5}$  with basis generators $X_i \equiv \{X_1, X_2, X_3, X_4, X_5\}$ and commutation table:
\begin{center}
	\begin{tabular}{| l | c|c |c|c| r| }
		\hline
		& $X_1$ & $X_2$ & $X_3$ & $X_4$ & $X_5$ \\ \hline
		$X_1$ & $0$ & 0 & 0 & 0 & 0\\ \hline
		$X_2$ & 0 & $0$ & 0  & $0$&$X_1$\\ \hline
		$X_3$ & 0 & 0 & $0$& 0 &$X_2$\\ \hline
		$X_4$ & 0 & 0 & 0 & $0$ &$X_3$\\ \hline
		$X_5$ & 0 & $-X_1$ &$-X_2$ & $-X_3$&$0$ \\ 
		\hline
	\end{tabular}
\end{center}
\noindent In this case $N(\mathfrak{g})=N_p(\mathfrak{g})=5-2=3$ and  $X_1$ is central. To obtain the other two polynomial Casimir invariants we need to find the other two independent solutions $F_{1,2}(x_1,x_2,x_3,x_4,x_5)$ of the following system of PDEs:
\begin{equation}
\begin{cases}
\hat{X_2}F&=x_2 F_{x_5}=0
\\
\hat{X_3}F&=x_3 F_{x_5} =0 
\\
\hat{X_4}F&=x_4 F_{x_5} =0 
\\
\hat{X_5}F&=-x_1 F_{x_2}-x_2 F_{x_3}-x_3 F_{x_4} =0 \, .
\end{cases}
\end{equation} 
They can be taken as \cite{sno14}:
\begin{equation}
\begin{cases}
F_1(x_1,x_2,x_3,x_4,x_5)=2 x_1 x_3- x_2^2 \\
F_2(x_1,x_2,x_3,x_4,,x_5) = x_2^3+3 x_1^2 x_4-3 x_1 x_2 x_3 \, .
\end{cases}
\label{[q:soln55}
\end{equation}
After symmetrization, and reordering terms, we get the following quadratic and cubic elements in $\mathcal{U}(\mathfrak{n}_{5,5})$:
\begin{equation}
C^{(1)}(X_1,X_2,X_3)=2X_1 X_3 - X_2^2 \, ,   \quad C^{(2)}(X_1,X_2,X_3,X_4)= X_2^3 + 3 X_1^2 X_4  - 3 X_1  X_2  X_3 \, .
\end{equation}
\noindent Following the same procedure as in the previous cases, we can now consider:
\begin{align}
&C_\alpha^{(1)}=2X_1^{[\alpha]} X_3^{[\alpha]} - \bigl(X_2^{[\alpha]}\bigl)^2  \\
&C_{\alpha \beta}^{(1)}= 2X_1^{[\alpha,\beta]} X_3^{[\alpha,\beta]} - \bigl(X_2^{[\alpha,\beta]}\bigl)^2\\
&C_{123}^{(1)}=2 X_1^{[1,2,3]} X_3^{[1,2,3]} - \bigl(X_2^{[1,2,3]}\bigl)^2 \, ,
\end{align}
\noindent and:
\begin{align}
&C_\alpha^{(2)}= \bigl(X_2^{[\alpha]}\bigl)^3 + 3 \bigl(X_1^{[\alpha]}\bigl)^2X_4^{[\alpha]}  - 3 X_1^{[\alpha]}  X_2^{[\alpha]}  X_3^{[\alpha]} \\
&C_{\alpha \beta}^{(2)}=  \bigl(X_2^{[\alpha ;\beta]}\bigl)^3 + 3 \bigl(X_1^{[\alpha]}\bigl)^2X_4^{[\alpha ;\beta]} - 3 X_1^{[\alpha ;\beta]}  X_2^{[\alpha ;\beta]}  X_3^{[\alpha ;\beta]}  \\
&C_{123}^{(2)}=  \bigl(X_2^{[1,2,3]}\bigl)^3 + 3 \bigl(X_1^{[1,2,3]}\bigl)^2X_4^{[1,2,3]}  - 3 X_1^{[1,2,3]} X_2^{[1,2,3]}  X_3^{[1,2,3]}  \, .
\end{align}
\noindent Thus, we again have two sets of polynomial Casimir invariants $\mathcal{C}^{(1)}$ and $\mathcal{C}^{(2)}$, but now one of them is composed by cubic elements. Again, the linear relation \eqref{eq:from} holds for the set $\mathcal{C}^{(1)}$ composed by quadratic elements.

\begin{remark}
	We observe that also the for the linear element $X_1$ a similar construction could be performed. However, in this case the elements $X_1^{[\alpha, \beta]}$ and $X_1^{[1,2,3]}$ would just result in a linear combination of the elements $X_1^{[\alpha]}$, that already commute with everything. Ideally, when this happens, we will be considering these central elements as \textquotedblleft commuting variables\textquotedblright, i.e. parameters.  
	\label{eq:rem}
\end{remark}
\noindent In this case, unlike the $\mathfrak{sl}(2)$ and $\mathfrak{so}(1,3)$ cases, there are no non-zero commutation relations among elements in the two sets. Everything commutes, then the algebra generated by $\{C_{12}^{(r)}, C_{23}^{(r)}, C_{13}^{(r)}\}$ is Abelian. Similar results can be obtained from many other Lie algebras, such as the Lie algebra $\mathfrak{n}_{6,7}$,  $\mathfrak{s}_{6,44}$ or $\mathfrak{s}_{6,92}$ just to mention a few. So, no further polynomial structures can be constructed from the Lie algebra $\mathfrak{n}_{5,5}$ with this approach.  This still allows to define algebraic Hamiltonian with a number of commuting polynomials in the enveloping algebra from the three copies. This fact is still interesting as the algebra can allow different realizations and the possibility of building integrable systems together with a set of commuting functions/operators. As an example in classical mechanics, where we consider the Poisson analogue of this Lie algebra to be realized in terms of canonical Poisson brackets in a given symplectic realization for the generators, we can consider the so-called $A_{5,2}$ (following the terminology used in \cite{doi:10.1063/1.522992}) integrable systems arising in the framework of coalgebra symmetry \cite{ballasco}, in our case with $N=3$. 

\subsubsection{On the use of subalgebras of $\mathfrak{n}_{5,5}$}
 \label{sec4.1.1}

\noindent Even if the construction applied to the Lie algebra $\mathfrak{n}_{5,5}$ leads to Abelian structures, it is still open the possibility of restricting to some of its subalgebras in order to obtain polynomial algebras in terms of elements defined in the enveloping algebra. As an explicit example, let us consider the following subalgebra of the Lie algebra $\mathfrak{n}_{5,5}$:
\begin{equation}
\mathfrak{n}_{5,5} \supset \hat{\mathfrak{g}}  := \text{span}\{Y_1, Y_2, Y_3\} \qquad  Y_1 := X_2\, ,\quad  Y_2 := X_5 \, , \quad Y_3 := X_1\, ,
\end{equation}
with commutation relations:
\begin{equation}
[Y_1, Y_2]=Y_3 \, , \quad [Y_1,Y_3]=[Y_2, Y_3]=0 \, ,
\label{commrels}
\end{equation}
\noindent and let us introduce the following elements in its enveloping algebra:
\begin{equation}
Z_1 := Y_1^2\, , \qquad Z_2 := Y_2^2 \, , \qquad Z_3 := Y_1 Y_2-Y_3/2 \, , \qquad Z_4:=Y_3 \, .
\label{eq:elements}
\end{equation}
These four elements satisfy the commutation relations:
\begin{equation}
[Z_1, Z_2]=4  Z_3 Z_4  \, , \quad [Z_1, Z_3]=2  Z_1 Z_4  \, , \quad [Z_2, Z_3]=-2 Z_2 Z_4  \, , \quad [Z_4, \cdot]=0 \, .
\label{comrels}
\end{equation}
\noindent Consider now the new rescaled generators $W_i:=Z_4^{-1}Z_i$ for $i=1,2,3$,  together with $W_4:=Z_4$, such that:
\begin{equation}
[W_1, W_2]=4 W_3 \, , \quad [W_1, W_3]=2  W_1 \, , \quad [W_2, W_3]=-2  W_2 \, , \quad [W_4, \cdot]=0 \, .
\label{comrelations}
\end{equation}
This algebra is endowed with the following Casimir operator:
\begin{equation}
C(W_1, W_2, W_3)= W_3^2-W_1W_2+2 W_3 \, .
\label{eq:ope}
\end{equation}
Let us now at this level consider three copies of it spanned by the generators $\{W_1^{[\alpha]},W_2^{[\alpha]}, W_3^{[\alpha]} , W^{[\alpha]}_4\}$ and the associated intermediate Casimir invariants:
\begin{align}
&C_\alpha^{(1)} =\bigl(W_3^{[\alpha]}\bigl)^2 - W_1^{[\alpha]}W_2^{[\alpha]}+2 W_3^{[\alpha]} \\
&C_{\alpha \beta}^{(1)}=\bigl(W_3^{[\alpha,\beta]}\bigl)^2 - W_1^{[\alpha,\beta]}W_2^{[\alpha,\beta]}+2 W_3^{[\alpha,\beta]}\\
&C_{123}^{(1)}=\bigl(W_3^{[1, 2, 3]}\bigl)^2 - W_1^{[1, 2, 3]}W_2^{[1, 2, 3]}+2 W_3^{[1, 2, 3]} \, .
\end{align}
At this point, let us re-express back the result in terms of the original generators $\{Z_1,Z_2,Z_3,Z_4\}$ and define the new polynomial elements:
\begin{equation}
\tilde{C}_\alpha^{(1)}:=\bigl(Z_4^{[\alpha]}\bigl)^2 C_\alpha^{(1)} =\bigl(Z_3^{[\alpha]}\bigl)^2-Z_1^{[\alpha]} Z_2^{[\alpha]}+2 Z_3^{[\alpha]}  Z_4^{[\alpha]} \, ,
\end{equation}
 \begin{align}
\tilde{C}_{\alpha \beta}^{(1)}:=\bigl(Z_4^{[\alpha]}\bigl)^2\bigl(Z_4^{[\beta]}\bigl)^2 C_{\alpha \beta}^{(1)}&=\bigl(Z_4^{[\alpha]}\bigl)^2 \bigl(Z_3^{[\beta]}\bigl)^2 - Z_1^{[\alpha]} Z_2^{[\alpha]} \bigl(Z_4^{[\alpha]}\bigl)^2 +2 Z_3^{[\alpha]}Z_4^{[\alpha]}  Z_3^{[\beta]}Z_4^{[\beta]} +\bigl(Z_3^{[\alpha]}\bigl)^2\bigl(Z_4^{[\beta]}\bigl)^2 \nonumber \\
&\hskip 0.25cm-Z_2^{[\alpha]}  Z_4^{[\alpha]} Z_1^{[\beta]} Z_4^{[\beta]}  -Z_1^{[\alpha]} Z_4^{[\alpha]} Z_2^{[\beta]}Z_4^{[\beta]}   -  Z_1^{[\alpha]} Z_2^{[\alpha]}\bigl(Z_4^{[\beta]}\bigl)^2 + Z_3^{[\alpha]}Z_4^{[\alpha]}\bigl(Z_4^{[\beta]}\bigl)^2 \nonumber \\
&\hskip 0.25cm+2 \bigl(Z_4^{[\alpha]}\bigl)^2 Z_3^{[\beta]} Z_4^{[\beta]} \, ,
\label{eq:polyele}
\end{align}
 \begin{align}
 \tilde{C}_{123}^{(1)}:=\bigl(Z_4^{[1]}\bigl)^2\bigl(Z_4^{[2]}\bigl)^2 \bigl(Z_4^{[3]}\bigl)^2 C_{123}^{(1)}&=(Z_4^{[3]})^2 \tilde{C}_{12}^{(1)}+(Z_4^{[1]})^2 \tilde{C}_{23}^{(1)}+(Z_4^{[2]})^2 \tilde{C}_{13}^{(1)} \nonumber \\
 &\hskip 0.25cm-(Z_4^{[2]})^2(Z_4^{[3]})^2\tilde{C}_1^{(1)}-(Z_4^{[1]})^2(Z_4^{[3]})^2\tilde{C}_2^{(1)}-(Z_4^{[1]})^2(Z_4^{[2]})^2\tilde{C}_3^{(1)} 
\label{eq:ali}
\end{align}
together with the generators:
\begin{align}
\tilde{Z}^{[1,2,3]}_i&:=Z_4^{[2]} Z_4^{[3]}Z_i^{[1]}+Z_4^{[1]} Z_4^{[3]}Z_i^{[2]}+Z_4^{[1]} Z_4^{[2]}Z_i^{[3]}  \quad (i=1,2,3)\\
\tilde{Z}^{[1,2,3]}_4 &=Z^{[1,2,3]}_4 \, .
\label{generators}
\end{align}

\noindent By direct computations, it is verified that these new polynomials satisfy the usual commutation relations:
 \begin{align}
[\tilde{C}^{[1]}_\alpha,\tilde{C}^{[1]}_{12}]&=[\tilde{C}^{[1]}_\alpha,\tilde{C}^{[1]}_{23}]=[\tilde{C}^{[1]}_\alpha,\tilde{C}^{[1]}_{13}]=0 \\
 [\tilde{C}^{[1]}_{123},\tilde{C}^{[1]}_{12}]&= [\tilde{C}^{[1]}_{123},\tilde{C}^{[1]}_{23}]= [\tilde{C}^{[1]}_{123},\tilde{C}^{[1]}_{13}]=0  \, ,
\label{eq:casimirtildae}
\end{align}
together with $[\tilde{C}^{[1]}_\alpha,\tilde{C}^{[1]}_{123}]=0$. Also, they commute with the elements $\tilde{Z}^{[1,2,3]}_k$, with $k=1,2,3,4$.

\noindent At this point, if we consider  the original subalgebra generators $\{Y_1, Y_2, Y_3\}$,  we see that the obtained reordered elements (notice that the central elements are taken to appear in front of each expression) are:

\begin{equation}
\bar{C}_\alpha^{(1)}=-\frac{3}{4}\bigl(Y_3^{[\alpha]}\bigl)^2 	
\label{c1}
\end{equation}
\begin{align}
\bar{C}_{\alpha \beta}^{(1)}=&-Y_3^{[\alpha]}Y_3^{[\beta]}\bigl(Y_2^{[\alpha]}\bigl)^2\bigl(Y_1^{[\beta]}\bigl)^2+2 Y_3^{[\alpha]}Y_3^{[\beta]} Y_1^{[\alpha]}Y_2^{[\alpha]}Y_1^{[\beta]}Y_2^{[\beta]}-Y_3^{[\alpha]}Y_3^{[\beta]}\bigl(Y_1^{[\alpha]}\bigl)^2\bigl(Y_2^{[\beta]}\bigl)^2 \nonumber \\
&-\bigl(Y_3^{[\alpha]}\bigl)^2Y_3^{[\beta]}Y_1^{[\beta]}Y_2^{[\beta]}  -Y_3^{[\alpha]}\bigl( Y_3^{[\beta]}\bigl)^2 Y_1^{[\alpha]}Y_2^{[\alpha]}-\bigl( Y_3^{[\alpha]}\bigl)^2\bigl( Y_3^{[\beta]}\bigl)^2
\label{cij}
\end{align}
\begin{align}
\bar{C}_{123}^{(1)}=-\frac{4}{3} \bar{C}_3^{(1)} \bar{C}_{12}^{(1)}-\frac{4}{3} \bar{C}_2^{(1)} \bar{C}_{13}^{(1)}-\frac{4}{3} \bar{C}_1^{(1)} \bar{C}_{23}^{(1)}-\frac{48}{9} \bar{C}_1^{(1)}\bar{C}_2^{(1)}\bar{C}_3^{(1)}  
\label{c123}
\end{align}
together with the generators:
\begin{align}
\bar{Z}^{[1,2,3]}_1&:=Y_3^{[2]} Y_3^{[3]}\bigl(Y_1^{[1]}\bigl)^2+Y_3^{[1]} Y_3^{[3]}\bigl(Y_1^{[2]}\bigl)^2+Y_3^{[1]} Y_3^{[2]}\bigl(Y_1^{[3]}\bigl)^2 \\
\bar{Z}^{[1,2,3]}_2&:=Y_3^{[2]} Y_3^{[3]}\bigl(Y_2^{[1]}\bigl)^2+Y_3^{[1]} Y_3^{[3]}\bigl(Y_2^{[2]}\bigl)^2+Y_3^{[1]} Y_3^{[2]}\bigl(Y_2^{[3]}\bigl)^2 \\
\bar{Z}^{[1,2,3]}_3&:=Y_3^{[2]} Y_3^{[3]}Y_1^{[1]}Y_2^{[1]}+Y_3^{[1]} Y_3^{[3]}Y_1^{[2]}Y_2^{[2]}+Y_3^{[1]} Y_3^{[2]}Y_1^{[3]}Y_2^{[3]}-\frac{3}{2}Y_3^{[1]}Y_3^{[2]}Y_3^{[3]}\\
\bar{Z}^{[1,2,3]}_4 &=Y_3^{[1]}+Y_3^{[2]} +Y_3^{[3]}  \, .
\label{generatorsb}
\end{align}

\noindent The set composed by the elements $\mathcal{\bar{C}}^{(1)}:=\{\bar{C}^{(1)}_1,\bar{C}^{(1)}_2,\bar{C}^{(1)}_3,\bar{C}^{(1)}_{12}, \bar{C}^{(1)}_{13}, \bar{C}^{(1)}_{23},\bar{C}^{(1)}_{123}\}$ can be now used to construct a polynomial algebra. In particular, we begin by defining the generators:
\begin{align}
	\bar{C}^{(1,1)}_{1223}&:= [\bar{C}^{(1)}_{12}, \bar{C}^{(1)}_{23}]=-\bar{C}^{(1,1)}_{2312}\\
	\bar{C}^{(1,1)}_{2313}&:=[\bar{C}^{(1)}_{23},\bar{C}^{(1)}_{13}]=-C^{(1,1)}_{1323}\\
	\bar{C}^{(1,1)}_{1312}&:=[\bar{C}^{(1)}_{13},\bar{C}^{(1)}_{12}]=-\bar{C}^{(1,1)}_{1213} \, .
	\label{g1}
	\end{align}
\noindent These elements are related as follows:
\begin{equation}
\bar{C}_1^{(1)}\bar{C}_3^{(1)} \bar{C}_{1223}^{(1,1)}=\bar{C}_1^{(1)}\bar{C}_2^{(1)} \bar{C}_{2313}^{(1,1)}=\bar{C}_2^{(1)}\bar{C}_3^{(1)} \bar{C}_{1312}^{(1,1)} \, .
\label{eq:rela}
\end{equation}
Thus, they commute each other. Moreover, together with the elements in the set $\mathcal{\bar{C}}^{(1)}$, they close in the following polynomial algebra:
\begin{align}
[\bar{C}^{(1)}_{12}, \bar{C}_{1223}^{(1,1)}]&=\frac{128}{9}\bigl(\bar{C}^{(1)}_{1} \bar{C}^{(1)}_{2} \bar{C}^{(1)}_{23} \bar{C}^{(1)}_{12}-\bigl(\bar{C}^{(1)}_{2}\bigl)^2  \bar{C}^{(1)}_{12} \bar{C}^{(1)}_{13}\bigl) \label{p1} \\
[\bar{C}^{(1)}_{23}, \bar{C}_{1223}^{(1,1)}]&=\frac{128}{9}\bigl(\bigl(\bar{C}^{(1)}_{2}\bigl)^2  \bar{C}^{(1)}_{13} \bar{C}^{(1)}_{23}-\bar{C}^{(1)}_{2} \bar{C}^{(1)}_{3}  \bar{C}^{(1)}_{23} \bar{C}^{(1)}_{12}\bigl) \label{p2} \\
[\bar{C}^{(1)}_{13}, \bar{C}_{1223}^{(1,1)}]&=\frac{128}{9}\bigl(\bar{C}^{(1)}_{2} \bar{C}^{(1)}_{3}  \bar{C}^{(1)}_{12} \bar{C}^{(1)}_{13}-\bar{C}^{(1)}_{1} \bar{C}^{(1)}_{2}  \bar{C}^{(1)}_{13} \bar{C}^{(1)}_{23}\bigl) \label{p3} \\
[\bar{C}^{(1)}_{12}, \bar{C}_{2313}^{(1,1)}]&=\frac{128}{9}\bigl(\bar{C}^{(1)}_{1} \bar{C}^{(1)}_{3}  \bar{C}^{(1)}_{23} \bar{C}^{(1)}_{12}-\bar{C}^{(1)}_{2} \bar{C}^{(1)}_{3}  \bar{C}^{(1)}_{12} \bar{C}^{(1)}_{13}\bigl) \label{p4} \\
[\bar{C}^{(1)}_{23}, \bar{C}_{2313}^{(1,1)}]&=\frac{128}{9}\bigl(\bar{C}^{(1)}_{2} \bar{C}^{(1)}_{3}  \bar{C}^{(1)}_{13} \bar{C}^{(1)}_{23}-\bigl(\bar{C}^{(1)}_{3}\bigl)^2 \bar{C}^{(1)}_{23} \bar{C}^{(1)}_{12}\bigl) \label{p5} \\
[\bar{C}^{(1)}_{13}, \bar{C}_{2313}^{(1,1)}]&=\frac{128}{9}\bigl(\bigl(\bar{C}^{(1)}_{3}\bigl)^2 \bar{C}^{(1)}_{12} \bar{C}^{(1)}_{13}-\bar{C}^{(1)}_{1} \bar{C}^{(1)}_{3}  \bar{C}^{(1)}_{13} \bar{C}^{(1)}_{23}\bigl) \label{p6} \\
 [\bar{C}^{(1)}_{12},\bar{C}_{1312}^{(1,1)}]&=\frac{128}{9}\bigl(\bigl(\bar{C}^{(1)}_{1}\bigl)^2 \bar{C}^{(1)}_{23} \bar{C}^{(1)}_{12}-\bar{C}^{(1)}_{1} \bar{C}^{(1)}_{2}  \bar{C}^{(1)}_{12} \bar{C}^{(1)}_{13}\bigl) \label{p7}  \\
 [\bar{C}^{(1)}_{23}, \bar{C}_{1312}^{(1,1)}]&=\frac{128}{9}\bigl(\bar{C}^{(1)}_{1} \bar{C}^{(1)}_{2} \bar{C}^{(1)}_{13} \bar{C}^{(1)}_{23}-\bar{C}^{(1)}_{1} \bar{C}^{(1)}_{3}  \bar{C}^{(1)}_{23} \bar{C}^{(1)}_{12}\bigl) \label{p8} \\
  [\bar{C}^{(1)}_{13},\bar{C}_{1312}^{(1,1)}]&=\frac{128}{9}\bigl(\bar{C}^{(1)}_{1} \bar{C}^{(1)}_{3} \bar{C}^{(1)}_{12} \bar{C}^{(1)}_{13}-\bigl(\bar{C}^{(1)}_{1}\bigl)^2\bar{C}^{(1)}_{13} \bar{C}^{(1)}_{23}\bigl) \, .\label{p9} 
\end{align}

\noindent Clearly, because of \eqref{eq:rela}, the following relations involving central elements also exist among those commutators:
\begin{align}
\bar{C}_1^{(1)}\bar{C}_3^{(1)}[\bar{C}^{(1)}_{12}, \bar{C}_{1223}^{(1,1)}]&=\bar{C}_1^{(1)}\bar{C}_2^{(1)}[\bar{C}^{(1)}_{12}, \bar{C}_{2313}^{(1,1)}] =\bar{C}_2^{(1)}\bar{C}_3^{(1)}[\bar{C}^{(1)}_{12}, \bar{C}_{1312}^{(1,1)}] \\
\bar{C}_1^{(1)}\bar{C}_3^{(3)}[\bar{C}^{(1)}_{23}, \bar{C}_{1223}^{(1,1)}]&=\bar{C}_1^{(1)}\bar{C}_2^{(1)}[\bar{C}^{(1)}_{23}, \bar{C}_{2313}^{(1,1)}] =\bar{C}_2^{(1)}\bar{C}_3^{(1)}[\bar{C}^{(1)}_{12}, \bar{C}_{2312}^{(1,1)}] \\
\bar{C}_1^{(1)}\bar{C}_3^{(1)}[\bar{C}^{(1)}_{12}, \bar{C}_{1323}^{(1,1)}]&=\bar{C}_1^{(1)}\bar{C}_2^{(1)}[\bar{C}^{(1)}_{13}, \bar{C}_{2313}^{(1,1)}] =\bar{C}_2^{(1)}\bar{C}_3^{(1)}[\bar{C}^{(1)}_{13}, \bar{C}_{1312}^{(1,1)}]  \, .
\label{eq:rels}
\end{align}

\noindent This should lead us to select only one independent element, say $\bar{C}_{1223}^{(1,1)}$ in \eqref{eq:rela}, and restrict to consider the relations involving commutators \eqref{p1}-\eqref{p3}. At that stage, we should also notice the following additional relation:
\begin{equation}
\bar{C}_3^{(1)}[\bar{C}^{(1)}_{12}, \bar{C}_{1312}^{(1,1)}] +\bar{C}_1^{(1)}[\bar{C}^{(1)}_{23}, \bar{C}_{1223}^{(1,1)}] +\bar{C}_2^{(1)}[\bar{C}^{(1)}_{13}, \bar{C}_{1223}^{(1,1)}] = 0 \, . 
\label{eq:addi}
\end{equation}

\noindent Thus, the generators of the polynomial algebra would be given by:
\begin{equation}
\{\bar{C}^{(1)}_{12}, \bar{C}^{(1)}_{23}, \bar{C}^{(1)}_{13}\} \cup \{\bar{C}^{(1,1)}_{1223}\} \, ,
\end{equation}
together with  the central elements:
\begin{equation}
\{Y_3^{[1]}, Y_3^{[2]}, Y_3^{[3]}\}  \cup \{\bar{C}_{123}^{(1)}\} \, ,
\label{eq:centralem}
\end{equation}
where we have taken into account the relation \eqref{c1}.

\subsection{ The nilpotent Lie algebra $\boldsymbol{\mathfrak{n}_{6,1}}$}
 \label{sec4.2}

\noindent  Let us consider the six-dimensional nilpotent Lie algebra $\mathfrak{n}_{6,1}$. This Lie algebra is endowed with the basis generators $X_i \equiv \{X_1, X_2, X_3, X_4, X_5, X_6\}$ and the commutation table reads:
\begin{center}
	\begin{tabular}{| l | c|c |c|c| c|r| }
		\hline
		& $X_1$ & $X_2$ & $X_3$ & $X_4$ & $X_5$ & $X_6$ \\ \hline
		$X_1$ & $0$ & $0$ &$0$ & $0$ & $0$& $0$\\ \hline
		$X_2$ & $0$ & $0$ & $0$  & $0$&$0$&$0$\\ \hline
		$X_3$ & $0$ & $0$ & $0$& $0$ &$0$&$0$\\ \hline
		$X_4$ & $0$ & $0$ &$0$ & $0$ &$X_2$&$X_3$\\ \hline
		$X_5$ & $0$ & $0$ &$0$ & $-X_2$&$0$ &$X_1$\\  \hline
		$X_6$ & $0$ & $0$ &$0$ & $-X_3$&$-X_1$ &$0$ \\
		\hline
	\end{tabular}
\end{center}
\noindent In this case $N(\mathfrak{g})=N_p(\mathfrak{g})=6-2=4$ and  $X_1, X_2, X_3$ are all central. To obtain the non-linear polynomial Casimir invariant we need to find the solution $F_{1}(x_1,x_2,x_3,x_4,x_5,x_6)$ of the following system of PDEs:
\begin{equation}
	\begin{cases}
		\hat{X_4}F&=x_2 F_{x_5}+x_3 F_{x_6} =0 
		\\
		\hat{X_5}F&=-x_2 F_{x_4}+x_1 F_{x_6}  =0 
		\\
		\hat{X_6}F&=-x_3 F_{x_4}-x_1 F_{x_5} =0 \, .
	\end{cases}
\end{equation} 
It turns out to be \cite{sno14}:
\begin{equation}
		F_1(x_1,x_2,x_3,x_4,x_5,x_6)=x_1 x_4+x_2 x_6-x_3 x_5 \, .
	\label{[sol5}
\end{equation}
After symmetrization and reordering terms we get the following quadratic element in $\mathcal{U}(\mathfrak{n}_{6,1})$:
\begin{equation}
	C^{(1)}(X_1,X_2,X_3,X_4,X_5,X_6)=X_1 X_4+X_2 X_6-X_3 X_5 \,  .
\end{equation}
\noindent Following the same procedure as in the previous cases, we  construct the following set of polynomial invariants:  
\begin{align}
	&C_\alpha^{(1)}=X_1^{[\alpha]} X_4^{[\alpha]} +X_2^{[\alpha]}  X_6^{[\alpha]} -X_3^{[\alpha]}  X_5^{[\alpha]}   \\
	&C_{\alpha \beta}^{(1)}= X_1^{[\alpha,\beta]} X_4^{[\alpha,\beta]}+X_2^{[\alpha,\beta]} X_6^{[\alpha,\beta]}-X_3^{[\alpha,\beta]} X_5^{[\alpha,\beta]}\\
	&C_{123}^{(1)}= X_1^{[1,2,3]} X_4^{[1,2,3]}+X_2^{[1,2,3]} X_6^{[1,2,3]}-X_3^{[1,2,3]}X_5^{[1,2,3]} \, .
\end{align}

\noindent  The  linear relation \eqref{eq:from} holds for the set $\mathcal{C}^{(1)}$ composed by quadratic elements. We introduce the new elements $\{C^{(1,1)}_{1223}, C^{(1,1)}_{2313}, C^{(1,1)}_{1312}\}$ as defined in \eqref{1a}-\eqref{1c} and, following the usual procedure we  find again $C_{1223}^{(1,1)}=C_{2313}^{(1,1)}=C_{1312}^{(1,1)}$.
\noindent Also, by direct computation we found that these commutators can be re-expressed just in terms of the central elements $\{X^{[\alpha]}_1 ,X^{[\alpha]}_2 ,X^{[\alpha]}_3 \} $ of the underlying Lie algebra. In fact, in this case, closure is achieved as:
\begin{align}
[C^{(1)}_{12}, C^{(1)}_{23}]=[C^{(1)}_{23}, C^{(1)}_{13}]=[C^{(1)}_{13}, C^{(1)}_{12}]&=X_1^{[1]} X_3^{[2]}  X_2^{[3]}- X_1^{[1]}X_2^{[2]}  X_3^{[3]} + X_2^{[1]}X_1^{[2]} X_3^{[3]} \nonumber \\
&\hskip 0.25cm-X_3^{[1]}X_1^{[2]}X_2^{[3]} +X_3^{[1]} X_2^{[2]} X_1^{[3]}   - X_2^{[1]}  X_3^{[2]}  X_1^{[3]}  \, .
\label{eq:commele}
\end{align}

\noindent Thus, the defining generators of the algebra are just the intermediate Casimir invariants:

\begin{equation}
\{C^{(1)}_{12}, C^{(1)}_{23}, C^{(1)}_{13} \} \, ,
\label{eq:int}
\end{equation}
together with the central elements:
\begin{equation}
\{X_1^{[1]}, X_1^{[2]},X_1^{[3]}, X_2^{[1]}, X_2^{[2]}, X_2^{[3]}, X_3^{[1]}, X_3^{[2]},X_3^{[3]} \} \cup \{C_1^{(1)}, C_2^{(1)}, C_3^{(1)}\} \cup \{C_{123}^{(1)}\} \, ,
\label{eq:centr}
\end{equation}
where we have taken into account the relation \eqref{eq:from}.
\begin{remark}
If we think in terms of applications to integrable/superintegrable systems, at a first glance this result seems to be not straightforward to understand as the algebra, even if not abelian, would just depend on central elements. To clarify this point, in the same spirit of \cite{ballasco}, we might think to realize this algebraic structure in a classical framework. One possible symplectic realization we found is the following:
\begin{align}
\xi_1&:=D(X_1)=\alpha_1  \, , \quad \xi_2:=D(X_2)=\alpha_2 \, , \quad \xi_3:=D(X_3)=\alpha_3    \\
 \xi_4&:=D(X_4)=\alpha_2 x \\ \xi_5&:=D(X_5)=p+\alpha_2 x+\alpha_2 \alpha_3 x^2 \\ \xi_6&:=D(X_6)=\frac{\alpha_3}{\alpha_2}p+(\alpha_3 -\alpha_1)x+\alpha_3^2 x^2 	\, ,
\label{symp}
\end{align}
with $\alpha_1, \alpha_3 \in \mathbb{R}$ and $\alpha_2 \in \mathbb{R}/\{0\}$. In this realization, when three copies are considered $($with a little abuse of notation$)$ we get:
\begin{align}
C_1^{(1)}&=C_2^{(1)}=C_3^{(1)}=0  \\
C_{12}^{(1)}&=(\alpha_3 \beta_2/\alpha_2-\beta_3)p_1+( \alpha_2\beta_3/\beta_2-\alpha_3)p_2+x_1(\alpha_2( \beta_1-\beta_3)+ \beta_2(\alpha_3-\alpha_1)+\alpha_3(\alpha_3 \beta_2-\alpha_2 \beta_3)x_1)\nonumber \\
&\hskip 0.25cm +(\alpha_2(\beta_3-\beta_1)+\beta_2(\alpha_1-\alpha_3))x_2+\beta_3 (\alpha_2 \beta_3-\alpha_3 \beta_2)x_2^2 \\
C_{13}^{(1)}&=(\alpha_3 \gamma_2/\alpha_2-\gamma_3)p_1+( \alpha_2\gamma_3/\gamma_2-\alpha_3)p_3+x_1(\alpha_2( \gamma_1-\gamma_3)+ \gamma_2(\alpha_3-\alpha_1)+\alpha_3(\alpha_3 \gamma_2-\alpha_2 \gamma_3)x_1)\nonumber \\
&\hskip 0.25cm +(\alpha_2(\gamma_3-\gamma_1)+\gamma_2(\alpha_1-\alpha_3))x_3+\gamma_3 (\alpha_2 \gamma_3-\alpha_3 \gamma_2)x_3^2 \\
C_{23}^{(1)}&=(\beta_3 \gamma_2/\beta_2-\gamma_3)p_2+( \beta_2\gamma_3/\gamma_2-\beta_3)p_3+x_2(\beta_2( \gamma_1-\gamma_3)+ \gamma_2(\beta_3-\beta_1)+\beta_3(\beta_3 \gamma_2-\beta_2 \gamma_3)x_2)\nonumber \\
&\hskip 0.25cm +(\beta_2(\gamma_3-\gamma_1)+\gamma_2(\beta_1-\alpha_3))x_3+\gamma_3 (\beta_2 \gamma_3-\beta_3 \gamma_2)x_3^2 \\
C_{123}^{(1)}&=-(\alpha_3+\gamma_3+\beta_3)(p_1+p_2+p_3+\alpha_2 x_1+\beta_2 x_2+\gamma_2 x_3+\alpha_2 \alpha_3 x_1^2+\beta_2 \beta_3 x_2^2+\gamma_2 \gamma_3 x_3^2) \nonumber \\
&\hskip 0.375cm +(\alpha_2+\beta_2+\gamma_2) \biggl(\frac{\alpha_3}{\alpha_2}p_1+\frac{\beta_3}{\beta_2}p_2+\frac{\gamma_3}{\gamma_2}p_3+(\alpha_3-\alpha_1)x_1+\alpha_3^2 x_1^2+(\beta_3-\beta_1)x_2+\beta_3^2 x_2^2 \nonumber\\
&\hskip 4cm+(\gamma_3-\gamma_1)x_3+\gamma_3^2 x_3^2\biggl) +(\alpha_1+\beta_1+\gamma_1)(\alpha_2 x_1+\beta_2 x_2+\gamma_2 x_3)
\label{eq:pcas}
\end{align}
where $\alpha_i,\beta_i,\gamma_i$, $($$i=1,2,3$$)$ are the central elements related to each copy, where we assume again $\alpha_2, \beta_2, \gamma_2 \in \mathbb{R}/\{0\}$. Thus, in principle, we have nine parameters to play with. It is immediate to verify that $C_{123}=C_{12}+C_{13}+C_{23}-C_1-C_2-C_3$ in the given symplectic realization. Also, the following Poisson commutation relations hold:
\begin{align}
\{C^{(1)}_{12}, C^{(1)}_{23}\}_{(\mathbf{q},\mathbf{p})}=\{C^{(1)}_{23}, C^{(1)}_{13}\}_{(\mathbf{q},\mathbf{p})}=\{C^{(1)}_{13}, C^{(1)}_{12}\}_{(\mathbf{q},\mathbf{p})}=\alpha_1 (\beta_3 \gamma_2- \beta_2 \gamma_3) &+\alpha_2 (\beta_1 \gamma_3-\beta_3\gamma_1) \nonumber \\
&+\alpha_3 (\gamma_1 \beta_2-\beta_1 \gamma_2) \, ,
\end{align}
where $\{f,g\}_{(\mathbf{q},\mathbf{p})}:=\sum_{i=1}^3 \bigl(\partial_{x_i}f \,\partial_{p_i} g -\partial_{x_i} g \,\partial_{p_i}f\bigl)$. Of course, we also have: 
\begin{equation}
\{C_{12}^{(1)},C_{123}^{(1)}\}_{(\mathbf{q},\mathbf{p})}=\{C_{23}^{(1)},C_{123}^{(1)}\}_{(\mathbf{q},\mathbf{p})}=\{C_{13}^{(1)},C_{123}^{(1)}\}_{(\mathbf{q},\mathbf{p})}=0 \, .
\end{equation}
\end{remark}
\subsection{ The nilpotent Lie algebra $\boldsymbol{\mathfrak{n}_{6,19}}$}
 \label{sec4.3}

\noindent  Let us consider the six-dimensional nilpotent Lie algebra $\mathfrak{n}_{6,19}$. This Lie algebra is endowed with the basis generators $X_i \equiv \{X_1, X_2, X_3, X_4, X_5, X_6\}$ and commutation table:
\begin{center}
	\begin{tabular}{| l | c|c |c|c| c|r| }
		\hline
		& $X_1$ & $X_2$ & $X_3$ & $X_4$ & $X_5$ & $X_6$ \\ \hline
		$X_1$ & $0$ & $0$ &$0$ & $0$ & $0$& $0$\\ \hline
		$X_2$ & $0$ & $0$ & $0$  & $0$&$0$&$X_1$\\ \hline
		$X_3$ & $0$ & $0$ & $0$& $X_1$ &$X_2$&$0$\\ \hline
		$X_4$ & $0$ & $0$ &$-X_1$ & $0$ &$X_3$&$X_2$\\ \hline
		$X_5$ & $0$ & $0$ &$-X_2$ & $-X_3$&$0$ &$X_4$\\  \hline
		$X_6$ & $0$ & $-X_1$ &$0$ & $-X_2$&$-X_4$ &$0$ \\
		\hline
	\end{tabular}
\end{center}
\noindent In this case $N(\mathfrak{g})=N_p(\mathfrak{g})=6-4=2$ and  $X_1$ is central. To obtain the other polynomial Casimir invariant we need to find the solution $F_{1}(x_1,x_2,x_3,x_4,x_5,x_6)$ of the following system of PDEs:
\begin{equation}
\begin{cases}
\hat{X_2}F&=x_1 F_{x_6}=0 
\\
\hat{X_3}F&=x_1 F_{x_4}+x_2 F_{x_5}  =0 
\\
\hat{X_4}F&=-x_1 F_{x_3}+x_3 F_{x_5}+x_2F_{x_6} =0\\
\hat{X_5}F&=-x_2 F_{x_3}-x_3 F_{x_4}+x_4F_{x_6} =0 \\
\hat{X_6}F&=-x_1 F_{x_2}-x_2 F_{x_4}-x_4F_{x_5} =0 \, .
\end{cases}
\end{equation} 
It reads \cite{sno14}:
\begin{equation}
F_1(x_1,x_2,x_3,x_4,x_5,x_6)=6x_1^2 x_5-6 x_1 x_2 x_4+3x_1 x_3^2+2 x_2^3 \, .
\label{sol}
\end{equation}
After symmetrization, and reordering terms, we get the following cubic element in $\mathcal{U}(\mathfrak{n}_{6,19})$:
\begin{equation}
C^{(1)}(X_1,X_2,X_3,X_4,X_5,X_6)=6X_1^2 X_5-6 X_1 X_2 X_4+3X_1 X_3^2+2 X_2^3  \,  .
\end{equation}
\noindent Again, we consider the following set of intermediate Casimir invariants:  
\begin{align}
&C_\alpha^{(1)}=6(X_1^{[\alpha]})^2 X_5^{[\alpha]}-6 X_1^{[\alpha]} X_2^{[\alpha]} X_4^{[\alpha]}+3X_1^{[\alpha]} (X_3^{[\alpha]})^2+2 (X_2^{[\alpha]})^3   \\
&C_{\alpha \beta}^{(1)}=6(X_1^{[\alpha,\beta]})^2 X_5^{[\alpha,\beta]}-6 X_1^{[\alpha,\beta]} X_2^{[\alpha,\beta]} X_4^{[\alpha,\beta]}+3X_1^{[\alpha,\beta]} (X_3^{[\alpha,\beta]})^2+2 (X_2^{[\alpha,\beta]})^3 \\
&C_{123}^{(1)}=6(X_1^{[1,2,3]})^2 X_5^{[1,2,3]}-6 X_1^{[1,2,3]} X_2^{[1,2,3]} X_4^{[1,2,3]}+3X_1^{[1,2,3]} (X_3^{[1,2,3]})^2+2 (X_2^{[1,2,3]})^3 \, ,
\end{align}
\noindent and the elements $\{C^{(1,1)}_{1223}, C^{(1,1)}_{2313}, C^{(1,1)}_{1312}\}$  as in \eqref{1a}-\eqref{1c}. Also in this case, they are not independent, since the following relation involving central elements holds:
\begin{equation}
(X_1^{[1]}+X_1^{[3]})C^{(1,1)}_{1223}=(X_1^{[1]}+X_1^{[2]})C^{(1,1)}_{2313}=(X_1^{[2]}+X_1^{[3]})C^{(1,1)}_{1312} \, .
\label{eq:c1}
\end{equation}
By taking commutators between these elements we get:
\begin{equation}
[C_{1223}^{(1,1)},C_{2313}^{(1,1)}]=[C_{2313}^{(1,1)},C_{1312}^{(1,1)}]=[C_{1312}^{(1,1)},C_{1223}^{(1,1)}]=0 \, .
\end{equation}
In this case, to reach closure, we need to consider higher order nested commutators. In particular,  we have to introduce the new elements:
\begin{align}
C^{(1,1,1)}_{121223}&:=[C^{(1)}_{12},[C^{(1)}_{12},C^{(1)}_{23}]] \, ,\quad C^{(1,1,1)}_{231223}:=[C^{(1)}_{23},[C^{(1)}_{12},C^{(1)}_{23}]]  \, ,\quad C^{(1,1,1)}_{131223}:=[C^{(1)}_{12},[C^{(1)}_{12},C^{(1)}_{23}]] \label{eq:r11}\\
C^{(1,1,1)}_{122313}&:=[C^{(1)}_{12},[C^{(1)}_{23},C^{(1)}_{13}]] \, ,\quad C^{(1,1,1)}_{232313}:=[C^{(1)}_{23},[C^{(1)}_{23},C^{(1)}_{13}]]  \, ,\quad C^{(1,1,1)}_{132313}:=[C^{(1)}_{13},[C^{(1)}_{23},C^{(1)}_{13}]] \\
C^{(1,1,1)}_{121312}&:=[C^{(1)}_{12},[C^{(1)}_{13},C^{(1)}_{12}]] \, ,\quad C^{(1,1,1)}_{231312}:=[C^{(1)}_{23},[C^{(1)}_{13},C^{(1)}_{12}]]  \, ,\quad C^{(1,1,1)}_{131312}:=[C^{(1)}_{13},[C^{(1)}_{13},C^{(1)}_{12}]]  \, .
\label{eq:newele}
\end{align}
Because of \eqref{eq:c1} we can restrict the analysis to the subset \eqref{eq:r11}. The three elements are not independent. In fact, the following relation involving central elements holds:
\begin{align}
(X_1^{[1]}+X_1^{[3]})(X_1^{[2]}+X_1^{[3]}) C^{(1,1,1)}_{121223}&+ (X_1^{[1]}+X_1^{[3]}) (X_1^{[1]}+X_1^{[2]}) C^{(1,1,1)}_{231223} \nonumber\\
&+ (X_1^{[1]}+X_1^{[2]}) (X_1^{[2]}+X_1^{[3]}) C^{(1,1,1)}_{131223}=0 \, .
\end{align}
Also, besides the fact that these three elements commute each other, i.e.:
\begin{equation}
[C^{(1,1,1)}_{121223},C^{(1,1,1)}_{231223}]=[C^{(1,1,1)}_{231223},C^{(1,1,1)}_{131223}]=[C^{(1,1,1)}_{131223},C^{(1,1,1)}_{121223}]=0 \, ,
\label{eq:comm}
\end{equation}
 the following commutation relations are also satisfied:
\begin{equation}
[C^{(1)}_{\alpha_1 \alpha_2},C^{(1,1,1)}_{\alpha_3 \alpha_4 \alpha_5 \alpha_6 \alpha_7 \alpha_8}]=0 \, , \quad[C^{(1,1)}_{\alpha_1 \alpha_2 \alpha_3 \alpha_4}, C_{\alpha_5 \alpha_6 \alpha_7 \alpha_8 \alpha_9 \alpha_{10}}^{(1,1,1)}]=0 \, .
\label{corel}
\end{equation}

\noindent In this case, what we get is again an Abelian structure, like for the five dimensional nilpotent Lie algebra $\mathfrak{n}_{5,5}$, but now obtained in terms of higher order nested commutators.

\section{Solvable Lie algebras}
\label{sec5}

This Section \ref{sec5} is devoted to the analysis of some representatives of \emph{solvable Lie algebras}. We recall that a Lie algebra is called \emph{solvable} if the derived  series terminates \cite{sno14}, i.e. if there exists $k \in  \mathbb{N}$ such that $g^{(k)} = 0$, where the derived series $\mathfrak{g}=\mathfrak{g}^{(0)} \supseteq \dots\supseteq  \mathfrak{g}^{(k)} \supseteq \dots $ is defined recursively: $\mathfrak{g}^{(k)}=[\mathfrak{g}^{(k-1)}, \mathfrak{g}^{(k-1)}]$, $\mathfrak{g}^{(0)}=\mathfrak{g}$.  

 \subsection{The solvable Lie algebra $\mathfrak{s}_{6,160}$}
  \label{sec5.1}

\noindent  Let us consider the six-dimensional Lie algebra $\mathfrak{s}_{6,160}$ in the basis generators $X_i \equiv \{X_1, X_2, X_3, X_4, X_5, X_6\}$ with commutation table: 

\begin{center}
	\begin{tabular}{| l | c|c |c|c| c|r| }
		\hline
		& $X_1$ & $X_2$ & $X_3$ & $X_4$ & $X_5$ & $X_6$ \\ \hline
		$X_1$ & $0$ & $0$ & $0$ & $0$ & $0$ & $0$\\ \hline
		$X_2$ & $0$ & $0$ & $0$  & $X_1$&$0$&$0$\\ \hline
		$X_3$ & $0$ & $0$ & $0$& $0$ &$X_1$&$-X_3$\\ \hline
		$X_4$ & $0$ & $-X_1$ &$0$ & $0$ &$0$&$-X_2$\\ \hline
		$X_5$ & $0$ & $0$ &$-X_1$ & $0$&$0$ &$X_5$\\  \hline
		$X_6$ & $0$ & $0$ &$X_3$ & $X_2$& $-X_5$ &$0$ \\
		\hline
	\end{tabular}
\end{center}
\noindent In this case $N(\mathfrak{g})=N_p(\mathfrak{g})=6-4=2$ and $X_1$ is central. To obtain the non-linear Casimir element we need to find the polynomial solution $F_{1}(x_1,x_2,x_3,x_4,x_5,x_6)$ of the following system of PDEs:
\begin{equation}
\begin{cases}
\hat{X_2}F&=x_1 F_{x_4} =0
\\
\hat{X_3}F&=x_1 F_{x_5}-x_3 F_{x_6} =0 
\\
\hat{X_4}F&=-x_1 F_{x_2} - x_2 F_{x_6}=0 
\\
\hat{X_5}F&=-x_1 F_{x_3}+x_5 F_{x_6}=0 
\\
\hat{X_6}F&=x_3 F_{x_3}+x_2 F_{x_4}-x_5 F_{x_5}=0 \, ,
\end{cases}
\end{equation} 
\noindent which is given by \cite{sno14}:
\begin{equation}
F_1(x_1,x_2,x_3,x_4,x_5,x_6)=2 x_1 x_6+2x_3 x_5-x_2^2 \, .
\label{cass}
\end{equation}
After symmetrization and reordering we obtain, discarding a term in $X_1$ (which is central), the following quadratic element in the enveloping algebra $\mathcal{U}(\mathfrak{s}_{6,160} )$:
\begin{equation}
C(X_1,X_2,X_3,X_4,X_5,X_6)=2X_1 X_6+2X_3 X_5 -  X_2^2 \, .
\label{eq:cas}
\end{equation}
\noindent Let us consider again three copies of the same Lie algebra, with generators $\{X_i^{[\alpha]}\}$ with $i=1,2,3,4,5,6$ and $\alpha=1,2,3$, respectively. From the above generators, we introduce the Casimir invariants:
\begin{align}
&C_\alpha^{(1)}=2X_1^{[\alpha]} X_6^{[\alpha]}+2X_3^{[\alpha]} X_5^{[\alpha]}- (X_2^{[\alpha]})^2 \\
&C_{\alpha \beta}^{(1)}=2X_1^{[\alpha,\beta]} X_6^{[\alpha,\beta]}+2X_3^{[\alpha,\beta]} X_5^{[\alpha,\beta]}-(X_2^{[\alpha,\beta]})^2\\
&C_{123}^{(1)}=2X_1^{[1,2,3]} X_6^{[1,2,3]}+2X_3^{[1,2,3]} X_5^{[1,2,3]}-(X_2^{[1,2,3]})^2 \, .
\end{align}
\noindent  Also in this case, the relation \eqref{eq:from} holds for these quadratic elements. 

\noindent The only non-zero commutation relations among the above elements are the ones among the two indices generators, and we use them to introduce the new generators:
$\{C^{(1,1)}_{1223}, C^{(1,1)}_{2313}, C^{(1,1)}_{1312}\}$
as in \eqref{1a}-\eqref{1c}.  Once again, they turn out to be not independent and, as a result, we are led to define the new element:
\begin{equation}
C_{1223}^{(1,1)}:=[C_{12}^{(1)},C^{(1)}_{23}]=[C^{(1)}_{23},C^{(1)}_{13}]=[C^{(1)}_{13},C^{(1)}_{12}] \, .
\label{eqs}
\end{equation}
Moreover, by considering higher order nested commutators we can then introduce the three elements:
\begin{equation}
C^{(1,1,1)}_{121223}=[C^{(1)}_{12},[C^{(1)}_{12},C^{(1)}_{23}]] \, ,\quad C^{(1,1,1)}_{231223}=[C^{(1)}_{23},[C^{(1)}_{12},C^{(1)}_{23}]]  \, ,\quad C^{(1,1,1)}_{131223}=[C^{(1)}_{12},[C^{(1)}_{12},C^{(1)}_{23}]] \, .
\label{eq:newelem}
\end{equation}
These elements are not independent, as the following linear relation holds:
\begin{equation}
 C^{(1,1,1)}_{121223}+C^{(1,1,1)}_{231223}+C^{(1,1,1)}_{131223} = 0 	 .
 \label{eq:linear}
 \end{equation}

\noindent In this case, by direct computations, we obtain that the three (dependent) elements \eqref{eq:linear}, together with lower-order ones, satisfy the following commutation relations involving the central elements $X_1^{[\alpha]}$ ($\alpha=1,2,3$):
\begin{equation}
\small{[C_{121223}^{(1,1,1)}, C_{231223}^{(1,1,1)}]=[C_{231223}^{(1,1,1)}, C_{131223}^{(1,1,1)}]=[C_{131223}^{(1,1,1)}, C_{121223}^{(1,1,1)}]=64X_1^{[1]} X_1^{[2]} X_1^{[3]} \bigl(X_1^{[1]} +X_1^{[2]} +X_1^{[3]} \bigl) C^{(1,1)}_{1223}} \, ,
\end{equation}
\begin{align}
[C^{(1)}_{12}, C_{121223}^{(1,1,1)}]&=4\bigl(X_1^{[1]}+X_1^{[2]}\bigl)^2 C_{1223}^{(1,1)} \\
[C^{(1)}_{23}, C_{121223}^{(1,1,1)}]&=4\bigl( X_1^{[1]} (X_1^{[3]} - X_1^{[2]}) - X_1^{[2]} (X_1^{[2]} + X_1^{[3]})\bigl) C_{1223}^{(1,1)} \\
[C^{(1)}_{13}, C_{121223}^{(1,1,1)}]&=4\bigl(X_1^{[3]}(X_1^{[2]} - X_1^{[1]})  - X_1^{[1]} (X_1^{[1]} + X_1^{[2]})\bigl) C_{1223}^{(1,1)} \\
[C^{(1)}_{12}, C_{231223}^{(1,1,1)}]&=4\bigl(X_1^{[1]} (X_1^{[3]}-X_1^{[2]}) - X_1^{[2]} (X_1^{[2]} + X_1^{[3]})\bigl) C_{1223}^{(1,1)} \\
[C^{(1)}_{23}, C_{231223}^{(1,1,1)}]&=4\bigl(X_1^{[2]}+X_1^{[3]}\bigl)^2 C_{1223}^{(1,1)} \\
[C^{(1)}_{13}, C_{231223}^{(1,1,1)}]&=4\bigl(X_1^{[1]} (X_1^{[2]} - X_1^{[3]}) - X_1^{[3]} (X_1^{[2]} + X_1^{[3]})) C_{1223}^{(1,1)} \\
[C^{(1)}_{12}, C_{131223}^{(1,1,1)}]&=4\bigl(X_1^{[3]} (X_1^{[2]} -X_1^{[1]} )- X_1^{[1]} (X_1^{[1]}  + X_1^{[2]})\bigl) C_{1223}^{(1,1)} \\
[C^{(1)}_{23}, C_{131223}^{(1,1,1)}]&=4\bigl(X_1^{[1]}  (X_1^{[2]}  - X_1^{[3]} ) - X_1^{[3]}  (X_1^{[2]}  + X_1^{[3]} )) C_{1223}^{(1,1)} \\
[C^{(1)}_{13}, C_{131223}^{(1,1,1)}]&=4(X_1^{[1]}  + X_1^{[3]} )^2 C_{1223}^{(1,1)}  \, ,
\label{eq:neweles}
\end{align}

\begin{align}
[C^{(1,1)}_{1223}, C_{121223}^{(1,1,1)}]&=4(X_1^{[1]}  (X_1^{[2]}  - X_1^{[3]} ) +  X_1^{[2]}  (X_1^{[2]}  + X_1^{[3]} )) C_{131223}^{(1,1,1)} \nonumber \\
&\hskip 0.5cm+4( X_1^{[3]}  (X_1^{[2]}  - X_1^{[1]} ) -  X_1^{[1]} (X_1^{[1]} + X_1^{[2]} )) C^{(1,1,1)}_{231223} \\
[C^{(1,1)}_{1223}, C_{231223}^{(1,1,1)}]&=4(X_1^{[1]}  (X_1^{[3]}  - X_1^{[2]} ) + X_1^{[3]}  (X_1^{[2]}  + X_1^{[3]} )) C_{121223}^{(1,1,1)} \nonumber \\
&\hskip 0.5cm+4( X_1^{[1]} (X_1^{[3]}  - X_1^{[2]} ) - X_1^{[2]}  (X_1^{[2]}  + X_1^{[3]} )) C^{(1,1,1)}_{131223} \\
[C^{(1,1)}_{1223}, C_{131223}^{(1,1,1)}]&=4(X_1^{[1]}  (X_1^{[2]}  - X_1^{[3]} ) - X_1^{[3]}  (X_1^{[2]}  + X_1^{[3]} )) C_{121223}^{(1,1,1)} \nonumber\\
& \hskip 0.5cm +4( X_1^{[1]}  (X_1^{[1]}  + X_1^{[2]} ) +  X_1^{[3]} (X_1^{[1]}  - X_1^{[2]} )) C^{(1,1,1)}_{231223} \, ,
\label{eq:newelesa}
\end{align}

\noindent with the additional constraints arising as a consequence of \eqref{eq:linear}: 
\begin{align}
[C^{(1)}_{12}, C_{121223}^{(1,1,1)}]+[C^{(1)}_{12}, C_{231223}^{(1,1,1)}]+[C^{(1)}_{13}, C_{131223}^{(1,1,1)}]&=0\\
[C^{(1)}_{23}, C_{121223}^{(1,1,1)}]+[C^{(1)}_{23}, C_{231223}^{(1,1,1)}]+[C^{(1)}_{23}, C_{131223}^{(1,1,1)}]&=0\\
[C^{(1)}_{13}, C_{121223}^{(1,1,1)}]+[C^{(1)}_{13}, C_{231223}^{(1,1,1)}]+[C^{(1)}_{13}, C_{131223}^{(1,1,1)}]&=0\\
[C^{(1,1)}_{1223}, C_{121223}^{(1,1,1)}]+[C^{(1,1)}_{1223}, C_{231223}^{(1,1,1)}]+[C^{(1,1)}_{1223}, C_{131223}^{(1,1,1)}]&= 0 \, .
\label{eq:relations}
\end{align}
\begin{remark}
\noindent Because of the linear relation \eqref{eq:linear} we might restrict to the commutation relations involving a subset of the two nested commutators composed by two linearly independent elements. In that case, our algebra generators could be taken as:
\begin{equation}
\{C_{12}^{(1)}, C_{23}^{(1)}, C_{13}^{(1)}\} \cup \{ C_{1223}^{(1,1)}\} \cup \{C_{121223}^{(1,1,1)}, C_{231223}^{(1,1,1)} \} \, ,
\label{eq:alg}
\end{equation}
together with the central elements:
\begin{equation}
\{X_1^{[1]}, X_1^{[2]}, X_1^{[3]} \} \cup \{C_1^{(1)}, C_2^{(1)}, C_3^{(1)}\} \cup \{C_{123}^{(1)} \} \, ,
\label{eq:central}
\end{equation}
taking into account also the usual linear relation among intermediate Casimir invariants \eqref{eq:from}.
\end{remark}
 \subsection{The solvable Lie algebra $\mathfrak{s}_{6,183}$}
  \label{sec5.2}
  
\noindent  Let us consider the six-dimensional Lie algebra $\mathfrak{s}_{6,183}$ with basis generators $X_i \equiv \{X_1, X_2, X_3, X_4, X_5, X_6\}$ and commutation table: 

\begin{center}
	\begin{tabular}{| l | c|c |c|c| c|r| }
		\hline
		& $X_1$ & $X_2$ & $X_3$ & $X_4$ & $X_5$ & $X_6$ \\ \hline
		$X_1$ & $0$ & $0$ & $0$ & $0$ & $0$ & $0$\\ \hline
		$X_2$ & $0$ & $0$ & $0$  & $0$&$X_1$&$X_2$\\ \hline
		$X_3$ & $0$ & $0$ & $0$& $X_1$ &$0$&$-2X_3$\\ \hline
		$X_4$ & $0$ & $0$ &$-X_1$ & $0$ &$X_2$&$2X_4$\\ \hline
		$X_5$ & $0$ & $-X_1$ &$0$ & $-X_2$&$0$ &$-X_5$\\  \hline
		$X_6$ & $0$ & $-X_2$ &$2X_3$ & $-2X_4$& $X_5$ &$0$ \\
		\hline
	\end{tabular}
\end{center}
\noindent In this case $N(\mathfrak{g})=N_p(\mathfrak{g})=6-4=2$ and $X_1$ is central. To obtain the cubic Casimir element we need to find the polynomial solution $F_{1}(x_1,x_2,x_3,x_4,x_5,x_6)$ of the following system of PDEs:
\begin{equation}
\begin{cases}
\hat{X_2}F&=x_1 F_{x_5}+x_2 F_{x_6} =0
\\
\hat{X_3}F&=x_1 F_{x_4}-2x_3 F_{x_6} =0 
\\
\hat{X_4}F&=-x_1 F_{x_3}+x_2 F_{x_5} +2 x_4 F_{x_6}=0 
\\
\hat{X_5}F&=-x_1 F_{x_2}-x_2 F_{x_4}-x_5 F_{x_6} =0 
\\
\hat{X_6}F&=-x_2 F_{x_2}+2x_3 F_{x_3}-2x_4F_{x_4}+x_5 F_{x_5}=0 \, .
\end{cases}
\end{equation} 
\noindent It turns out to be \cite{sno14}:
\begin{equation}
F_1(x_1,x_2,x_3,x_4,x_5,x_6)=x_1^2 x_6+ 2x_1x_3 x_4- x_1 x_2 x_5-x_2^2 x_3 \, .
\label{[q:solss}
\end{equation}
After symmetrization, reordering the obtained terms and discarding a term proportional to $X_1^2$ we get the following cubic element in the enveloping algebra $\mathcal{U}(\mathfrak{s}_{6,183} )$:
\begin{equation*}
C(X_1,X_2,X_3,X_4,X_5,X_6)=X_1^2 X_6+2X_1 X_3 X_4- X_1 X_2 X_5  - X_2^2 X_3\, .
\label{eq:commrels}
\end{equation*}
\noindent Let us consider now three copies of the same solvable Lie algebra with generators $\{X_i^{[\alpha]}\}$,  $i=1,2,3,4,5,6$ and $\alpha=1,2,3$ respectively. Consider then the new elements:
\begin{align}
&C_\alpha=\bigl(X_1^{[\alpha]}\bigl)^2 X_6^{[\alpha]}+2X_1^{[\alpha]} X_3^{[\alpha]} X_4^{[\alpha]}- X_1^{[\alpha]} X_2^{[\alpha]} X_5^{[\alpha]}  - \bigl(X_2^{[\alpha]}\bigl)^2 X_3^{[\alpha]} \\
&C_{\alpha \beta}=\bigl(X_1^{[\alpha, \beta]}\bigl)^2 X_6^{[\alpha, \beta]}+2X_1^{[\alpha, \beta]} X_3^{[\alpha, \beta]} X_4^{[\alpha, \beta]}- X_1^{[\alpha, \beta]} X_2^{[\alpha, \beta]} X_5^{[\alpha, \beta]}  - \bigl(X_2^{[\alpha, \beta]}\bigl)^2X_3^{[\alpha, \beta]}\\
&C_{123}=\bigl(X_1^{[1,2,3]}\bigl)^2 X_6^{[1,2,3]}+2X_1^{[1,2,3]} X_3^{[1,2,3]} X_4^{[1,2,3]}- X_1^{[1,2,3]} X_2^{[1,2,3]} X_5^{[1,2,3]}  - \bigl(X_2^{[1,2,3]}\bigl)^2X_3^{[1,2,3]} \, .
\end{align}

\noindent Again, the only non-zero commutation relations among the elements of the set of these polynomial Casimir invariants are the ones involving the two indices generators, and we use them to introduce the new generators $\{C^{(1,1)}_{1223}, C^{(1,1)}_{2313}, C^{(1,1)}_{1312}\}$ as in \eqref{1a}-\eqref{1c}. Then, by considering higher order nested commutators we can introduce the additional elements:
\begin{align}
C^{(1,1,1)}_{121223}&=[C^{(1)}_{12},[C^{(1)}_{12},C^{(1)}_{23}]] \, ,\quad C^{(1,1,1)}_{231223}=[C^{(1)}_{23},[C^{(1)}_{12},C^{(1)}_{23}]]  \, ,\quad C^{(1,1,1)}_{131223}=[C^{(1)}_{12},[C^{(1)}_{12},C^{(1)}_{23}]] \label{eq:r1}\\
C^{(1,1,1)}_{122313}&=[C^{(1)}_{12},[C^{(1)}_{23},C^{(1)}_{13}]] \, ,\quad C^{(1,1,1)}_{232313}=[C^{(1)}_{23},[C^{(1)}_{23},C^{(1)}_{13}]]  \, ,\quad C^{(1,1,1)}_{132313}=[C^{(1)}_{13},[C^{(1)}_{23},C^{(1)}_{13}]] \label{eq:r2} \\
C^{(1,1,1)}_{121312}&=[C^{(1)}_{12},[C^{(1)}_{13},C^{(1)}_{12}]] \, ,\quad C^{(1,1,1)}_{231312}=[C^{(1)}_{23},[C^{(1)}_{13},C^{(1)}_{12}]]  \, ,\quad C^{(1,1,1)}_{131312}=[C^{(1)}_{13},[C^{(1)}_{13},C^{(1)}_{12}]]  \, .
\label{eq:r3}
\end{align}
\noindent Together with the relation $C^{(1,1,1)}_{131223} + C^{(1,1,1)}_{122313} + C_{231312}^{(1,1,1)}=0$, there exist linear relations involving central elements among these nested commutators, i.e.:
\begin{align}
&(X_1^{[1]} + X_1^{[3]} )^2 (X_1^{[2]}  + X_1^{[3]} ) C_{121223}^{(1,1,1)} + (X_1^{[1]} + 
X_1^{[2]}) (X_1^{[1]} + X_1^{[3]}) (X_1^{[2]} + X_1^{[3]}) C_{131223}^{(1,1,1)} \nonumber\\
- &(X_1^{[2]} + X_1^{[3]})^2 (X_1^{[1]} + X_1^{[2]}) C_{131312}^{(1,1,1)}- (X_1^{[1]} + X_1^{[2]}) (X_1^{[2]} + X_1^{[3]}) (X_1^{[1]} + X_1^{[3]}) C_{231312}^{(1,1,1)}   \nonumber \\
+ &(X_1^{[1]} + X_1^{[3]})^2 (X_1^{[1]} + X_1^{[2]}) C_{231223}^{(1,1,1)}   
- (X_1^{[2]} + X_1^{[3]})^2 (X_1^{[1]} + 
X_1^{[3]})  C_{121312}^{(1,1,1)}  =0 \, , \label{eq:rel2}\\
& (X_1^{[1]} + X_1^{[2]})^2 (X_1^{[1]} + 
X_1^{[3]})  C_{232313}^{(1,1,1)}- (X_1^{[2]} + X_1^{[3]}) (X_1^{[1]} + X_1^{[2]}) (X_1^{[1]} + X_1^{[3]}) C_{131223}^{(1,1,1)}  \nonumber\\
  - &(X_1^{[1]} + X_1^{[2]}) (X_1^{[2]} + X_1^{[3]})^2 C^{(1,1,1)}_{131312}- 2 (X_1^{[1]} + X_1^{[2]}) (X_1^{[2]} + X_1^{[3]})  (X_1^{[1]} + X_1^{[3]}) C^{(1,1,1)}_{231312} \nonumber\\
  +&(X_1^{[1]} + X_1^{[2]})^2 (X_1^{[2]} + X_1^{[3]}) C_{132313}^{(1,1,1)}
  - (X_1^{[2]} + X_1^{[3]})^2 (X_1^{[1]} + X_1^{[2]}) C_{121312}^{(1,1,1)} =0 \, .\label{eq:rel3} 
\end{align}

\noindent In this case, closure among higher order nested commutators is more involved but still can be obtained. 


\noindent In order to describe how closure is reached in this case, we proceed formally  by introducing the following sets of generators:
\begin{equation}
\mathcal{N}_0:=\{C_{12}^{(1)},C_{23}^{(1)}, C_{13}^{(1)}\} \, ,  \,\,  \mathcal{N}_1:=\{[\mathcal{N}_0, \mathcal{N}_0]\} \,, \,\,  \mathcal{N}_2:=\{[\mathcal{N}_0, \mathcal{N}_{1}]\} \, , \,\,\dots\,\, \, ,\,\, \mathcal{N}_{k^*}:=\{[\mathcal{N}_0, \mathcal{N}_{k^*-1}]\} 
\label{formaset}
\end{equation}
where $k^*$ is obtained by computing all possible commutation relations among the defining generators in the sets until closure is achieved. Here, we indicated generators formally, and with the sets we mean that we have to consider all nested commutation relations in a given set. For example, the set $\mathcal{N}_1$ has to be understood as a set composed by the three generators $\{C^{(1,1)}_{1223}, C^{(1,1)}_{2313}, C^{(1,1)}_{1312}\}$, the set $\mathcal{N}_2$ as the one composed by the generators \eqref{eq:r1}-\eqref{eq:r3} and so on. Again, we remark that as soon as a new set of generators is introduced, one has to check for the existence of linear relations (also involving central elements) among them, such as for example  \eqref{eq:rel2}-\eqref{eq:rel3} for the elements in the set $\mathcal{N}_2$. Due to the cumbersome expressions we have obtained for formulas related to this Lie algebra, and considering the actual number of elements required to close, we will not be presenting explicitly all relations among all higher order nested commutators, rather we just focus on explaining how higher order nested commutators close among each other. 

\noindent  We have been able to verify computationally that closure is achieved for $k^*=4$, so that it is possible to close all commutation relations in terms of the elements in the set:
\begin{equation}
\mathcal{N}:=\mathcal{N}_0 \cup \mathcal{N}_1 \cup \mathcal{N}_2  \cup \mathcal{N}_3 \cup \mathcal{N}_4 	\, .
\label{sets}
\end{equation}

\noindent In particular, we realized that the obtained relations can be presented formally as:

\begin{equation}
P_{a,b}\bigl(X_1^{[1]},X_1^{[2]},X_1^{[3]}\bigl)[\mathcal{N}_a, \mathcal{N}_b] = \sum_{c=1}^4 Q_{a,b}^c \bigl(X_1^{[1]},X_1^{[2]},X_1^{[3]}\bigl) \mathcal{N}_c  \label{eqf}
\end{equation}
or, more explicitly:
\begin{align}
P_{1,1}\bigl(X_1^{[1]},X_1^{[2]},X_1^{[3]}\bigl)[\mathcal{N}_1, \mathcal{N}_1] &= Q_{1,1}^1\bigl(X_1^{[1]},X_1^{[2]},X_1^{[3]}\bigl) \mathcal{N}_1+Q_{1,1}^2\bigl(X_1^{[1]},X_1^{[2]},X_1^{[3]}\bigl)\mathcal{N}_2 \label{eqf1}\\
P_{1,2}\bigl(X_1^{[1]},X_1^{[2]},X_1^{[3]}\bigl)[\mathcal{N}_1, \mathcal{N}_2] &= Q_{1,2}^4\bigl(X_1^{[1]},X_1^{[2]},X_1^{[3]}\bigl) \mathcal{N}_4 \label{eqf2}\\
P_{1,3}\bigl(X_1^{[1]},X_1^{[2]},X_1^{[3]}\bigl)[\mathcal{N}_1, \mathcal{N}_3] &=Q_{1,3}^1 \bigl(X_1^{[1]},X_1^{[2]},X_1^{[3]}\bigl) \mathcal{N}_1+Q_{1,3}^2 \bigl(X_1^{[1]},X_1^{[2]},X_1^{[3]}\bigl) \mathcal{N}_2  \label{eqf3}\\
P_{1,4}\bigl(X_1^{[1]},X_1^{[2]},X_1^{[3]}\bigl)[\mathcal{N}_1, \mathcal{N}_4] &=Q_{1,4}^2 \bigl(X_1^{[1]},X_1^{[2]},X_1^{[3]}\bigl) \mathcal{N}_2+Q_{1,4}^4 \bigl(X_1^{[1]},X_1^{[2]},X_1^{[3]}\bigl) \mathcal{N}_4 \label{eqf4} \\
P_{2,2}\bigl(X_1^{[1]},X_1^{[2]},X_1^{[3]}\bigl)[\mathcal{N}_2, \mathcal{N}_2] &=Q_{2,2}^1\bigl(X_1^{[1]},X_1^{[2]},X_1^{[3]}\bigl) \mathcal{N}_1+Q_{2,2}^3\bigl(X_1^{[1]},X_1^{[2]},X_1^{[3]}\bigl)\mathcal{N}_3      \label{eqf5}\\
P_{2,3}\bigl(X_1^{[1]},X_1^{[2]},X_1^{[3]}\bigl)[\mathcal{N}_2, \mathcal{N}_3] &=Q_{2,3}^2\bigl(X_1^{[1]},X_1^{[2]},X_1^{[3]}\bigl) \mathcal{N}_2+Q_{2,3}^4\bigl(X_1^{[1]},X_1^{[2]},X_1^{[3]}\bigl) \mathcal{N}_4 \label{eqf6} \\
P_{2,4}\bigl(X_1^{[1]},X_1^{[2]},X_1^{[3]}\bigl)[\mathcal{N}_2, \mathcal{N}_4] &=Q_{2,4}^1\bigl(X_1^{[1]},X_1^{[2]},X_1^{[3]}\bigl) \mathcal{N}_1+Q_{2,4}^3\bigl(X_1^{[1]},X_1^{[2]},X_1^{[3]}\bigl)\mathcal{N}_3  \label{eqf8}\\
P_{3,3}\bigl(X_1^{[1]},X_1^{[2]},X_1^{[3]}\bigl)[\mathcal{N}_3, \mathcal{N}_3] &=Q_{3,3}^1\bigl(X_1^{[1]},X_1^{[2]},X_1^{[3]}\bigl) \mathcal{N}_1+Q_{3,3}^2\bigl(X_1^{[1]},X_1^{[2]},X_1^{[3]}\bigl)\mathcal{N}_2  \label{eqf7} \\
P_{3,4}\bigl(X_1^{[1]},X_1^{[2]},X_1^{[3]}\bigl)[\mathcal{N}_3, \mathcal{N}_4] &=Q_{3,4}^2\bigl(X_1^{[1]},X_1^{[2]},X_1^{[3]}\bigl) \mathcal{N}_2+Q_{3,4}^4\bigl(X_1^{[1]},X_1^{[2]},X_1^{[3]}\bigl)\mathcal{N}_4  \label{eqf9}\\
P_{4,4}\bigl(X_1^{[1]},X_1^{[2]},X_1^{[3]}\bigl)[\mathcal{N}_4, \mathcal{N}_4] &=  Q_{4,4}^1\bigl(X_1^{[1]},X_1^{[2]},X_1^{[3]}\bigl) \mathcal{N}_1+Q_{4,4}^3\bigl(X_1^{[1]},X_1^{[2]},X_1^{[3]}\bigl)\mathcal{N}_3 \label{eqf10}
\end{align}

\noindent These commutation relations have to be understood in the following way. On the left hand side the commutators are computed among all the possible generators in a given set $\mathcal{N}_i$ and $\mathcal{N}_j$, $i,j=1,2,3,4$. The right hand side indicates that those commutation relations among nested commutators can be then re-expressed as linear combinations of terms belonging to the sets $\mathcal{N}_i$ ($i=1, 2,3,4$) if polynomials involving central elements appear in both sides of the equations as multiplicative factors. As per our convention, the right hand side has to be understood as a sum of elements in the sets, multiplied by some specific polynomial functions of the central elements. For example, if we focus on the sets $\mathcal{N}_1$ and $\mathcal{N}_2$, whose cardinality is $\text{card}(\mathcal{N}_1)=3$ and $\text{card}(\mathcal{N}_2)=9$  respectively, and we indicate the elements in $\mathcal{N}_0$, whose cardinality is $\text{card}(\mathcal{N}_0)=3$, as:
\begin{equation}
\mathcal{N}_0^{\{1\}}:=C_{12}^{(1)} \qquad \mathcal{N}_0^{\{2\}}:=C_{23}^{(1)}  \qquad \mathcal{N}_0^{\{3\}}:=C_{13}^{(1)} 
\label{no}
\end{equation}
then, elements in the first set can be indicated as:
\begin{equation}
\mathcal{N}_1^{\{1\}}:=[\mathcal{N}_0^{\{1\}}, \mathcal{N}_0^{\{2\}} ] \qquad \mathcal{N}_1^{\{2\}}:=[\mathcal{N}_0^{\{2\}}, \mathcal{N}_0^{\{3\}} ]  \qquad \mathcal{N}_1^{\{3\}}:=[\mathcal{N}_0^{\{3\}}, \mathcal{N}_0^{\{1\}} ] 
\label{eq:ele}
\end{equation}
whereas in the second one as:
\begin{equation}
\mathcal{N}_2^{\{i,j\}}:=[\mathcal{N}_0^{\{i\}},\mathcal{N}_1^{\{j\}} ]\, , \quad  i=1, \dots ,\text{card}(\mathcal{N}_0) \, ,\quad j=1, \dots, \text{card}(\mathcal{N}_1) \, .
\label{eq:elem}
\end{equation}
Notice that in this case $\text{card}(\mathcal{N}_1)=\text{card}(\mathcal{N}_0)=3$, as there not exist linear relations among elements in the set $\mathcal{N}_1$. 
That said, the first relation \eqref{eqf1} is in fact a formal expression representing a set of relations among elements \eqref{eq:ele}, \eqref{eq:elem} and the commutators:
\begin{equation}
[\mathcal{N}_1^{\{i\}},\mathcal{N}_1^{\{j\}}] \qquad \bigl(i,j=1, \dots, \text{card}(\mathcal{N}_1)\bigl) \, .
\label{eq:commn1n1}
\end{equation}

\noindent  To be more specific, the first relation indicates in a formal way that elements \eqref{eq:commn1n1}, taken as:
\begin{equation}
P_{1,1}^{\{i,j\}}[\mathcal{N}_1^{\{i\}},\mathcal{N}_1^{\{j\}}] \quad \bigl(i,j =1, \dots \text{card}(\mathcal{N}_1)\bigl) \, ,
\label{eq:setlhs}
\end{equation}
where $P_{1,1}^{\{i,j\}}$ at fixed index $i=i^*,j=j^*$ is a polynomial function of the central elements, can be re-expressed as linear expansions (involving polynomials of the central elements) of the generators appearing in the sets $\mathcal{N}_1$ and $\mathcal{N}_2$, i.e:
\begin{equation}
 \sum_{k=1}^{\text{card}(\mathcal{N}_1)} \bigl(Q^{1,i,j}_{1,1} \bigl)^{\{k\}}\mathcal{N}_1^{\{k\}} +  \sum_{k=1}^{\text{card}(\mathcal{N}_0)}\sum_{\ell=1}^{\text{card}(\mathcal{N}_1)} \bigl(Q^{2,i,j}_{1,1} \bigl)^{\{k,\ell\}} \mathcal{N}_2^{\{k,\ell\}} \, .
\label{eq:setrhs}
\end{equation}
Thus,  \eqref{eqf1} is in fact a collection of  $\text{card}(\mathcal{N}_1) \times \text{card}(\mathcal{N}_1)=9$ relations of the type\footnote{Clearly, in this case, when $i=j$  the elements are the same. Thus, the number of relations restricts to  $\text{card}(\mathcal{N}_1) \times \text{card}(\mathcal{N}_1)-3=6$, and in turn to the $3$ independent ones (the relations among the elements \eqref{eq:ele}). }:
{\footnotesize \begin{equation}
P_{1,1}^{\{i,j\}}[\mathcal{N}_1^{\{i\}},\mathcal{N}_1^{\{j\}}]= \sum_{k=1}^{\text{card}(\mathcal{N}_1)} \bigl(Q^{1,i,j}_{1,1} \bigl)^{\{k\}}\mathcal{N}_1^{\{k\}} +  \sum_{k=1}^{\text{card}(\mathcal{N}_0)}\sum_{\ell=1}^{\text{card}(\mathcal{N}_1)} \bigl(Q^{2,i,j}_{1,1} \bigl)^{\{k,\ell\}} \mathcal{N}_2^{\{k,\ell\}}\, ,  \quad \bigl(i,j =1, \dots \text{card}(\mathcal{N}_1)\bigl) \, ,
\label{eq:reinvcent}
\end{equation}}

\noindent where $Q_{1,1}^{1,i,j}$ and $Q_{1,1}^{2,i,j}$ at fixed indices $i=i^*, j=j^*$ might be understood as a vector and a matrix respectively,  their components   $\bigl(Q_{1,1}^{1,i^*,j^*}\bigl)^{\{k\}}$, $\bigl(Q_{1,1}^{2,i^*,j^*}\bigl)^{\{k,\ell\}}$ being polynomial functions of the central elements $X_1^{[\alpha]}$, $\alpha=1,2,3$. Let us consider an explicit example to clarify these points. If we take $i=1$ and $j=2$, the relation that we obtain is the following:
\begin{align}
P_{1,1}^{\{1,2\}}[\mathcal{N}_1^{\{1\}}, \mathcal{N}_1^{\{2\}}] =& \,\bigl(Q_{1,1}^{1,1,2}\bigl)^{\{1\}} \mathcal{N}_1^{\{1\}}+\bigl(Q_{1,1}^{1,1,2}\bigl)^{\{2\}} \mathcal{N}_1^{\{2\}}+\bigl(Q_{1,1}^{1,1,2}\bigl)^{\{3\}} \mathcal{N}_1^{\{3\}}+\bigl(Q_{1,1}^{2,1,2}\bigl)^{\{1,1\}}\mathcal{N}_2^{\{1,1\}}\nonumber\\
&+\bigl(Q_{1,1}^{2,1,2}\bigl)^{\{1,3\}}\mathcal{N}_2^{\{1,3\}}+\bigl(Q_{1,1}^{2,1,2}\bigl)^{\{3,2\}}\mathcal{N}_2^{\{3,2\}}+\bigl(Q_{1,1}^{2,1,2}\bigl)^{\{3,3\}}\mathcal{N}_2^{\{3,3\}}
\label{eq:relfix12}
\end{align}
with: 
\begin{equation}
P_{1,1}^{\{1,2\}}(X_1^{[1]},X_1^{[2]},X_1^{[3]})=\bigl(X_1^{[1]}+X_1^{[2]}\bigl)^2\bigl(X_1^{[1]}+X_3^{[1]}\bigl)^2 \, ,
\label{eq:rel}
\end{equation}
\begin{align}
\bigl(Q_{1,1}^{1,1,2}\bigl)^{\{1\}}&=2 \bigl(X_1^{[1]}+X_1^{[2]}\bigl)^2\bigl(X_1^{[2]}+X_1^{[3]}\bigl)\bigl(X_1^{[3]}+X_1^{[1]}\bigl)^3 \bigl(X_1^{[1]}\bigl(X_1^{[3]}-X_1^{[2]}\bigl)+X_1^{[3]}\bigl(X_1^{[2]}+X_1^{[3]}\bigl)\bigl)\\
\bigl(Q_{1,1}^{1,1,2}\bigl)^{\{2\}}&=2 \bigl(X_1^{[1]} + X_1^{[2]} \bigl)^3 \bigl(X_1^{[1]}  + X_1^{[3]} \bigl)^2 \bigl(X_1^{[2]}  + X_1^{[3]}\bigl) \bigl(X_1^{[1]}  \bigl(X_1^{[2]}  - X_1^{[3]} \bigl) + 
X_1^{[2]} \bigl (X_1^{[2]}  + X_1^{[3]}\bigl)\bigl)\\
\bigl(Q_{1,1}^{1,1,2}\bigl)^{\{3\}}&=-2 \bigl(X_1^{[1]} + X_1^{[2]}\bigl)^2 \bigl(X_1^{[1]} + X_1^{[3]}\bigl)^2 \bigl(X_1^{[2]} + X_1^{[3]}\bigl)^4
\label{eq:al}
\end{align}
and:
\begin{align}
	\bigl(Q_{1,1}^{2,1,2}\bigl)^{\{1,1\}}&=\bigl(X_1^{[1]} + X_1^{[3]} \bigl)^4 \bigl(X_1^{[2]}  + X_1^{[3]} \bigl)^2\\
\bigl(Q_{1,1}^{2,1,2}\bigl)^{\{1,3\}}&=-\bigl(X_1^{[1]} + X^{[3]} \bigl)^3 (X_1^{[2]}  + X_1^{[3]})^3\\
\bigl(Q_{1,1}^{2,1,2}\bigl)^{\{3,2\}}&=\bigl(X_1^{[1]} + X_1^{[2]} \bigl)^4 \bigl(X_1^{[2]}  + X_1^{[3]} \bigl)^2\\
		\bigl(Q_{1,1}^{2,1,2}\bigl)^{\{3,3\}}&=-\bigl(X_1^{[1]} + X_1^{[2]} \bigl)^3 \bigl(X_1^{[2]}  + X_1^{[3]} \bigl)^3 \, ,
	\label{eq:al212}
	\end{align}

\noindent This means that the commutator $\bigl[[C^{(1)}_{12}, C^{(1)}_{23}], [C^{(1)}_{23}, C^{(1)}_{13}]\bigl]$ can be re-expressed in terms of the nested commutators $[C^{(1)}_{12}, C^{(1)}_{23}]$, $[C^{(1)}_{23}, C^{(1)}_{13}]$, $[C^{(1)}_{13}, C^{(1)}_{12}]$, $[C^{(1)}_{12}, [C^{(1)}_{12}, C^{(1)}_{23}]]$, $[C^{(1)}_{12}, [C^{(1)}_{13}, C^{(1)}_{12}]]$, $[C^{(1)}_{13}, [C^{(1)}_{23}, C^{(1)}_{13}]]$ and $[C^{(1)}_{13}, [C^{(1)}_{13}, C^{(1)}_{12}]]$ modulo the appearance of polynomial functions of central elements in both sides of the equation.
Another explicit example, for $i=1$, $j=3$, is the relation:
\begin{align}
P_{1,1}^{\{1,3\}}[\mathcal{N}_1^{\{1\}}, \mathcal{N}_1^{\{3\}}] =& \,\bigl(Q_{1,1}^{1,1,3}\bigl)^{\{1\}} \mathcal{N}_1^{\{1\}}+\bigl(Q_{1,1}^{1,1,3}\bigl)^{\{2\}} \mathcal{N}_1^{\{2\}}+\bigl(Q_{1,1}^{1,1,3}\bigl)^{\{1\}} \mathcal{N}_1^{\{3\}}+\bigl(Q_{1,1}^{2,1,3}\bigl)^{\{2,1\}}\mathcal{N}_2^{\{2,1\}}\nonumber \\
&+\bigl(Q_{1,1}^{2,1,3}\bigl)^{\{2,2\}}\mathcal{N}_2^{\{2,2\}}
+\bigl(Q_{1,1}^{2,1,3}\bigl)^{\{3,2\}}\mathcal{N}_2^{\{3,2\}}+\bigl(Q_{1,1}^{2,1,3}\bigl)^{\{3,3\}}\mathcal{N}_2^{\{3,3\}} \, ,
\label{eq:relfix13}
\end{align}
\noindent with: 
\begin{equation}
P_{1,1}^{\{1,3\}}(X_1^{[1]},X_1^{[2]},X_1^{[3]})=\bigl(X_1^{[1]}+X_1^{[3]}\bigl)^2\bigl(X_1^{[2]}+X_3^{[3]}\bigl)^2 \, ,
\label{eq:rel13}
\end{equation}
\begin{align}
\bigl(Q_{1,1}^{1,1,3}\bigl)^{\{1\}}&=2 \bigl(X_1^{[1]} + X_1^{[2]}\bigl) \bigl(X_1^{[1]} + X_1^{[3]}\bigl)^3 (X_1^{[2]} + X_1^{[3]}\bigl)^2 \bigl(X_1^{[2]} X_1^{[3]} -X_1^{[1]}\bigl(X_1^{[1]}+X_1^{[2]}+X_1^{[3]}\bigl) \bigl)\\
\bigl(Q_{1,1}^{1,1,3}\bigl)^{\{2\}}&=2 \bigl(X_1^{[1]} + X_1^{[2]} \bigl)^4 \bigl(X_1^{[1]}  + X_1^{[3]} \bigl)^2 \bigl(X_1^{[2]}  + X_1^{[3]} \bigl)^2\\
	\bigl(Q_{1,1}^{1,1,3}\bigl)^{\{3\}}&=-2 \bigl(X_1^{[1]} + X_1^{[2]}\bigl) \bigl(X_1^{[1]} + X_1^{[3]}\bigl)^2 \bigl(X_1^{[2]} + X_1^{[3]}\bigl)^3 \bigl(X_1^{[1]}\bigl (X_1^{[2]} - X_1^{[3]}\bigl) + 
	X_1^{[2]} \bigl(X_1^{[2]} + X_1^{[3]}\bigl)\bigl)
	\label{eq:al2}
	\end{align}
and:
\begin{align}
\bigl(Q_{1,1}^{2,1,3}\bigl)^{\{2,1\}}&=-\bigl(X_1^{[1]} + X_1^{[2]}\bigl)^2 \bigl(X_1^{[1]} + X_1^{[3]}\bigl)^4\\
	\bigl(Q_{1,1}^{2,1,3}\bigl)^{\{2,2\}}&=\bigl(X_1^{[1]} + X_1^{[2]}\bigl)^3 \bigl(X_1^{[1]} + X_1^{[3]}\bigl)^3\\
\bigl(Q_{1,1}^{2,1,3}\bigl)^{\{3,2\}}&=\bigl(X_1^{[1]} + X_1^{[2]} \bigl)^3 \bigl(X_1^{[2]}  + X_1^{[3]} \bigl)^3\\
\bigl(Q_{1,1}^{2,1,3}\bigl)^{\{3,3\}}&=-\bigl(X_1^{[1]}  + X_1^{[2]} \bigl)^2 \bigl(X_1^{[2]}  + X_1^{[3]} \bigl)^4\, .
\label{eq:al3}
\end{align}

\noindent The same reasoning applies for all other relations involving higher order nested commutators. 

 To summarize, starting from the set $\mathcal{N}_0$ composed by intermediate Casimir invariants, closure is achieved as follows:
\begin{align}
[\mathcal{N}_0,\mathcal{N}_{i-1}]&=:\mathcal{N}_i \qquad i=1,2,3,4 \label{geni}\\
P_{a,b}\bigl(X_1^{[1]},X_1^{[2]},X_1^{[3]}\bigl)[\mathcal{N}_a, \mathcal{N}_b] &= \sum_{c=1}^4 Q_{a,b}^c \bigl(X_1^{[1]},X_1^{[2]},X_1^{[3]}\bigl) \mathcal{N}_c
\label{eq:closureachieved}
\end{align}
with $Q_{a,b}^c$ as in \eqref{eqf1}-\eqref{eqf10}. Finally, we notice that besides the relations \eqref{geni}, the additional higher order nested commutators $[\mathcal{N}_0, \mathcal{N}_4]$ have to be considered. It turns out that they are all expressible in terms of lower order ones through the following (formal) relation involving central elements:
\begin{equation}
P_{04}\bigl(X_1^{[1]},X_1^{[2]},X_1^{[3]}\bigl)[\mathcal{N}_0,\mathcal{N}_4] = Q_{04}^1\bigl(X_1^{[1]},X_1^{[2]},X_1^{[3]}\bigl)\mathcal{N}_1+Q_{04}^3\bigl(X_1^{[1]},X_1^{[2]},X_1^{[3]}\bigl)\mathcal{N}_3 \, .
\label{eq:expr}
\end{equation}
For example, a relatively simple relation of this type is given by:
\begin{align}
[C_{13}^{(1)},[ C_{13}^{(1)}, [C_{13}^{(1)}, [C_{13}^{(1)}, [C_{13}^{(1)}, C_{12}^{(1)}]]]]] = 
&-4  \bigl(X_1^{[1]} + X_1^{[3]}\bigl)^8 [C^{(1)}_{13}, C^{(1)}_{12}] \nonumber \\
&+5 \bigl(X_1^{[1]} + X_1^{[3]}\bigl)^4 [C^{(1)}_{13}, [C^{(1)}_{13}, [C^{(1)}_{13}, C^{(1)}_{12}]]]\, .
\label{eq:fiven}
\end{align}
\section{Levi decomposable Lie algebras}
\label{sec6}
This final Section \ref{sec6} is devoted to the analysis of some representatives of \emph{Levi decomposable Lie algebras}. We recall that Levi decomposable Lie algebras are nonsolvable Lie algebras $\mathfrak{g}$ with nonvanishing nilradical $\mathfrak{n}$ (maximal nilpotent ideal of $\mathfrak{g}$) such as $\mathfrak{g}=\mathfrak{p} \niplus \mathfrak{r}$, $[\mathfrak{r},\mathfrak{g}] \subseteq \mathfrak{n} \subseteq \mathfrak{r}$, $[\mathfrak{p},\mathfrak{p}]=\mathfrak{p}$, where $\mathfrak{p}$ is a semisimimple Lie algebra (the Levi factor) and $\mathfrak{r}$ is the radical (maximal solvable ideal of $\mathfrak{g}$) \cite{sno14}.
 \subsection{The Levi decomposable Lie algebra $\mathfrak{sl}_2 \niplus 3 \mathfrak{n}_{1,1}$}
\label{sec6.1}

\noindent  Let us consider the six-dimensional Levi decomposable Lie algebra $\mathfrak{sl}_2 \niplus 3 \mathfrak{n}_{1,1}$ with basis generators $X_i \equiv \{X_1, X_2, X_3, X_4, X_5, X_6\}$ and commutation table:

\begin{center}
	\begin{tabular}{| l | c|c |c|c| c|r| }
		\hline
		& $X_1$ & $X_2$ & $X_3$ & $X_4$ & $X_5$ & $X_6$ \\ \hline
		$X_1$ & $0$ & 2$X_1$ &$-X_2$ & $0$ & $2X_4$& $-X_5$\\ \hline
		$X_2$ & $-2X_1$ & $0$ & $2X_3$  & $-2X_4$&$0$&$2X_6$\\ \hline
		$X_3$ & $X_2$ & $-2X_3$ & $0$& $X_5$ &$-2X_6$&$0$\\ \hline
		$X_4$ & $0$ & $2X_4$ &$-X_5$ & $0$ &$0$&$0$\\ \hline
		$X_5$ & $-2X_4$ & $0$ &$2X_6$ & $0$&$0$ &$0$\\  \hline
		$X_6$ & $X_5$ & $-2X_6$ &$0$ & $0$&$0$ &$0$ \\
		\hline
	\end{tabular}
\end{center}
\noindent In this case $N(\mathfrak{g})=N_p(\mathfrak{g})=6-4=2$ and to obtain the quadratic Casimir elements we need to find the two independent solutions $F_{1,2}(x_1,x_2,x_3,x_4,x_5,x_6)$ of the following system of PDEs:
\begin{equation}
\begin{cases}
\hat{X_1}F&=2x_1 F_{x_2}-x_2 F_{x_3}+2x_4 F_{x_5}-x_5 F_{x_6} =0
\\
\hat{X_2}F&=-2x_1 F_{x_1}+2x_3 F_{x_3}-2x_4 F_{x_4}+2x_6 F_{x_6} =0
\\
\hat{X_3}F&=x_2 F_{x_1}-2x_3 F_{x_2}+x_5 F_{x_4}-2x_6 F_{x_5} =0 
\\
\hat{X_4}F&=2x_4 F_{x_2}-x_5 F_{x_3} =0 
\\
\hat{X_5}F&=-2x_4 F_{x_1}+2x_6 F_{x_3} =0 
\\
\hat{X_6}F&=x_5 F_{x_1}-2x_6 F_{x_2}=0 \, . 
\end{cases}
\end{equation} 
\noindent The two solutions are given by \cite{sno14}:
\begin{equation}
\begin{cases}
F_1(x_1,x_2,x_3,x_4,x_5,x_6)=2x_1 x_6+ x_2 x_5+2 x_3 x_4 \\
F_2(x_1,x_2,x_3,x_4,x_5,x_6) = x_5^2+4 x_4x_6 \, .
\end{cases}
\label{[q:sols}
\end{equation}
After symmetrization and reordering we then obtain the following quadratic elements in $\mathcal{U}(\mathfrak{sl}_2(\mathbb{F}) \niplus 3 \mathfrak{n}_{1,1} )$:
\begin{equation}
C^{(1)}(X_1,X_2,X_3,X_4,X_5,X_6)=2X_1 X_6+X_2 X_5+2X_3X_4 \qquad C^{(2)}(X_4,X_5,X_6)=X_5^2+ 4 X_4 X_6 \, .
\label{eq:commrel}
\end{equation}
\noindent Let us consider again three copies of the same Lie algebra, with generators $\{X_i^{[\alpha]}\}$, $i=1,2,3,4,5,6$ and $\alpha=1,2,3$  respectively. From the above generators, we introduce the Casimir invariants:
\begin{align}
&C_\alpha^{(1)}=2X_1^{[\alpha]} X_6^{[\alpha]}+X_2^{[\alpha]} X_5^{[\alpha]}+2 X_3^{[\alpha]}X_4^{[\alpha]} \\
&C_{\alpha \beta}^{(1)}=2X_1^{[\alpha,\beta]} X_6^{[\alpha,\beta]}+X_2^{[\alpha,\beta]} X_5^{[\alpha,\beta]}+2X_3^{[\alpha,\beta]}X_4^{[\alpha,\beta]}\\
&C_{123}^{(1)}=2X_1^{[1,2,3]} X_6^{[1,2,3]}+X_2^{[1,2,3]} X_5^{[1,2,3]}+2X_3^{[1,2,3]}X_4^{[1,2,3]}
\end{align}
\noindent and:
\begin{align}
&C_\alpha^{(2)}=\bigl(X_5^{[\alpha]}\bigl)^2+4 X_4^{[\alpha]} X_6^{[\alpha]} \\
&C_{\alpha \beta}^{(2)}=\bigl(X_5^{[\alpha \beta]}\bigl)^2+4X_4^{[\alpha,\beta]} X_6^{[\alpha \beta]}\\
&C_{123}^{(2)}=\bigl(X_5^{[1,2,3]}\bigl)^2+4 X_4^{[1,2,3]} X_6^{[1,2,3]} \, .
\end{align}
\noindent  Again, the relations \eqref{linrelr} hold for both quadratic elements in the two sets $\mathcal{C}^{(r)}$. The only non-zero commutation relations among the elements of the set $\mathcal{C}^{(1)}$ are the ones among the two indices generators, and we use them to introduce the  elements $\{C^{(1,1)}_{1223}, C^{(1,1)}_{2313}, C^{(1,1)}_{1312}\}$ as reported in \eqref{1a}-\eqref{1c}.

\noindent Again, they are not independent. In particular, by proceeding as in the previous cases, we got the constraint $a_3=-a_1-a_2$, which lead us to define the additional element:
\begin{equation}
C_{1223}^{(1,1)}:=[C_{12}^{(1)},C^{(1)}_{23}]=[C^{(1)}_{23},C^{(1)}_{13}]=[C^{(1)}_{13},C^{(1)}_{12}] \, .
\label{eq}
\end{equation}

\noindent Among the elements of the second set $\mathcal{C}^{(2)}=\{C^{(2)}_1,C^{(2)}_2,C^{(2)}_3,C^{(2)}_{12},C^{(2)}_{13},C^{(2)}_{23},C^{(2)}_{123}\}$ there are no non-zero commutation relations. Among mixed elements from both of the two sets there are non-zero commutation relations, and we use them to introduce the new elements:
\begin{align}
C^{(1,2)}_{1223}&:= [C^{(1)}_{12}, C^{(2)}_{23}]=-C^{(2,1)}_{2312}  \qquad C^{(2,1)}_{1223}:= [ C^{(2)}_{12}, C^{(1)}_{23}]=-C^{(1,2)}_{2312} \\
C^{(1,2)}_{2313}&:=[C^{(1)}_{23},C^{(2)}_{13}]=-C^{(2,1)}_{1323}\qquad C^{(2,1)}_{2313}:= [ C^{(2)}_{23}, C^{(1)}_{13}]=-C^{(1,2)}_{1323}\\
C^{(1,2)}_{1312}&:=[C^{(1)}_{13},C^{(2)}_{12}]=-C^{(2,1)}_{1213} \qquad C^{(2,1)}_{1312}:= [ C^{(2)}_{13}, C^{(1)}_{12}]=-C^{(1,2)}_{1213}  \, .
\end{align}
These elements turn out to be not independent and, again, lead us to define the additional generator:
\begin{equation}
C_{1223}^{(1,2)}:=[C^{(1)}_{12}, C^{(2)}_{23}]=[C^{(1)}_{23}, C^{(2)}_{13}]=[C^{(1)}_{13}, C^{(2)}_{12}]=[C^{(2)}_{12}, C^{(1)}_{23}]=[C^{(2)}_{23}, C^{(1)}_{13}]=[C^{(2)}_{13}, C^{(1)}_{12}] \, .
\end{equation}
If we now compute the relations among these elements, besides $[C_{1223}^{(1,1)}, C_{1223}^{(1,2)}]=0$, we get:
\begin{align}
 [C_{12}^{(1)},C^{(1,1)}_{1223}]&=2\bigl(C^{(1)}_{12}(C^{(2)}_{23}-C^{(2)}_{13})+(C^{(1)}_{23}-C^{(1)}_{13})C^{(2)}_{12}\bigl)\nonumber\\
&\hskip 0.25cm+2\bigl((C^{(1)}_2-C^{(1)}_1)(C^{(2)}_3-C^{(2)}_{123})+(C^{(1)}_3-C^{(1)}_{123})(C^{(2)}_2-C^{(2)}_1)\bigl)\\
 [C_{23}^{(1)},C^{(1,1)}_{1223}]&=2\bigl(C^{(1)}_{23}(C^{(2)}_{13}-C^{(2)}_{12})+(C^{(1)}_{13}-C^{(1)}_{12})C^{(2)}_{23}\bigl)\\
 &\hskip 0.25cm+2\bigl((C^{(1)}_3-C^{(1)}_2)(C^{(2)}_1-C^{(2)}_{123})+(C^{(1)}_1-C^{(1)}_{123})(C^{(2)}_3-C^{(2)}_2)\bigl)\\
 [C_{13}^{(1)},C^{(1,1)}_{1223}]&=2\bigl(C^{(1)}_{13}(C^{(2)}_{12}-C^{(2)}_{23})+(C^{(1)}_{12}-C^{(1)}_{23})C^{(2)}_{13}\bigl)\\
& \hskip 0.25cm+2\bigl((C^{(1)}_1-C^{(1)}_3)(C^{(2)}_2-C^{(2)}_{123})+(C^{(1)}_2-C^{(1)}_{123})(C^{(2)}_1-C^{(2)}_3)\bigl) 	\, ,
\end{align}
\begin{align}
 [C_{12}^{(2)},C^{(1,1)}_{1223}]&=2(C^{(2)}_{23}C^{(2)}_{12}-C^{(2)}_{12}C^{(2)}_{13}+(C^{(2)}_2-C^{(2)}_1)(C^{(2)}_3-C^{(2)}_{123}))\\
[C_{23}^{(2)},C^{(1,1)}_{1223}]&=2(C^{(2)}_{13} C^{(2)}_{23}-C^{(2)}_{23}C^{(2)}_{12}+(C^{(2)}_3-C^{(2)}_2)(C^{(2)}_1-C^{(2)}_{123}))\\
 [C_{13}^{(2)},C^{(1,1)}_{1223}]&=2(C^{(2)}_{12} C^{(2)}_{13}-C^{(2)}_{13}C^{(2)}_{23}+(C^{(2)}_1-C^{(2)}_3)(C^{(2)}_2-C^{(2)}_{123})) \, ,
\end{align}
\begin{align}
[C_{12}^{(1)},C^{(1,2)}_{1223}]&= [C_{12}^{(2)},C^{(1,1)}_{1223}]\hskip 0.525cm
[C_{23}^{(1)},C^{(1,2)}_{1223}]= [C_{23}^{(2)},C^{(1,1)}_{1223}]\hskip 0.525cm
[C_{13}^{(1)},C^{(1,2)}_{1223}]=  [C_{13}^{(2)},C^{(1,1)}_{1223}] \\
[C_{12}^{(2)},C^{(1,2)}_{1223}]&=0\hskip 2.2275cm [C_{23}^{(2)},C^{(1,2)}_{1223}]=0
\hskip 2.2275cm
[C_{13}^{(2)},C^{(1,2)}_{1223}]=0
\end{align}
with:
$$ [C_{12}^{(1)},C^{(1,1)}_{1223}]+ [C_{23}^{(1)},C^{(1,1)}_{1223}]+ [C_{13}^{(1)},C^{(1,1)}_{1223}]=0 \, , \quad [C_{12}^{(1)},C^{(1,2)}_{1223}]+[C_{23}^{(1)},C^{(1,2)}_{1223}]+[C_{13}^{(1)},C^{(1,2)}_{1223}]=0 \, .$$

\noindent Thus, also for this Levi decomposable Lie algebra, our construction lead to a quadratic algebra closed in terms of elements belonging to both of the two sets of polynomial Casimir invariants. 
 \subsection{The Levi decomposable Lie algebra $\mathfrak{sl}_2 \niplus \mathfrak{n}_{3,1}$: a virtual copies related approach}
 \label{sec6.2}
 
\noindent  The main aim of this brief final Section \ref{sec6.2}  is to point out how virtual copies of the Levi factor of a Levi decomposable Lie algebra might be used as a tool to construct \emph{virtual copies of polynomial algebras} within our approach. To this aim, let us consider the Levi decomposable Lie algebra $\mathfrak{sl}_2 \niplus \mathfrak{n}_{3,1}$. It is the semidirect sum of a Levi factor $\mathfrak{sl}_2$ and a radical given by the Heisenberg algebra $\mathfrak{n}_{3,1}$. This non-semisimple Lie algebra arises in many physical problems, within our context we mention that has been investigated recently from the point of view of commutants of algebraic Hamiltonians \cite{CAMPOAMORSTURSBERG2022168694}. Here, following \cite{sno14}, we are using a different (relabelled and rescaled) basis. In a classical (Poisson) setting it appears in the framework of coalgebra symmetry in relation to the two-photon Lie-Poisson (co)algebra \cite{ballesteros2001two, Ballesteros_2009}. 

\noindent The Levi decomposable Lie algebra $\mathfrak{sl}_2 \niplus  \mathfrak{n}_{3,1}$ has six basis generators $X_i \equiv \{X_1, X_2, X_3, X_4, X_5, X_6\}$ and commutation table: 

\begin{center}
	\begin{tabular}{| l | c|c |c|c| c|r| }
		\hline
		& $X_1$ & $X_2$ & $X_3$ & $X_4$ & $X_5$ & $X_6$ \\ \hline
		$X_1$ & $0$ & $2X_1$ & $-X_2$ & $0$ & $X_6$ & $0$\\ \hline
		$X_2$ & $-2X_1$ & $0$ & $2X_3$  & $0$&$X_5$&$-X_6$\\ \hline
		$X_3$ & $X_2$ & $-2X_3$ & $0$& $0$ &$0$&$X_5$\\ \hline
		$X_4$ & $0$ & $0$ &$0$ & $0$ &$0$&$0$\\ \hline
		$X_5$ & $-X_6$ & $-X_5$ &$0$ & $0$&$0$ &$X_4$\\  \hline
		$X_6$ & $0$ & $X_6$ &$-X_5$ & $0$& $-X_4$ &$0$ \\
		\hline
	\end{tabular}
\end{center}

\noindent The element $X_4$ is central, and there is a cubic Casimir invariant \cite{sno14}:
\begin{equation}
C(X_1,X_2,X_3,X_4,X_5,X_6)=4 X_1 X_3 X_4-2 X_1 X_5^2+X_2^2X_4 + 2X_2X_5X_6 + 2X_3X_6^2+X_2 X_4 \, .
\label{Casc}
\end{equation}
\noindent Let us now define the generators of the radical as:
\begin{equation}
Y_1:=X_4 \, , \quad Y_2:=X_5, \, \quad  Y_3:=X_6 
\label{genrad} 
\end{equation}
\noindent and let us consider the following elements in the enveloping algebra $\mathcal{U}(\mathfrak{sl}_2 \niplus  \mathfrak{n}_{3,1})$:
\begin{equation}
X_1':=Y_1 X_1+Y_3^2/2 \, , \quad X_2':=Y_1 X_2+Y_2 Y_3-Y_1/2\, , \quad X_3'=Y_1 X_3-Y_2^2/2  \, .
\label{envalg}
\end{equation}
These elements, together with $Y_1$, define a \textquotedblleft virtual copy\textquotedblright of the Levi factor \cite{cam09}:
\begin{equation}
[Y_1, X_i']=0 \qquad [X'_i, X'_j]=Y_1C_{ij}^kX'_k \, ,
\label{eq:vcop}
\end{equation}
where $C_{ij}^k$ are the structure constants related to the Levi factor $\mathfrak{sl}_2$ generated by $\{X_1, X_2, X_3\}$. Explicitly:
\begin{equation}
[Y_1, X_i']=0 \, ,\quad [X'_1, X'_2]=2Y_1X'_1 \, ,\quad [X'_1, X'_3]=-Y_1X'_2 \, , \quad [X'_2, X'_3]=2Y_1X'_3 \, .
\label{eq:vcopr}
\end{equation}
Once a virtual copy has been constructed, as a result of the method, from the Casimir of the Levi factor:
\begin{equation}
 C_{\mathfrak{sl}_2}(X_1,X_2,X_3)=X_2^2+2(X_1 X_3+X_3 X_1)=X_2^2+4 X_1 X_3+2X_2
 \label{eq:caslevi}
 \end{equation}
  we can obtain the Casimir \eqref{Casc} by substituting the generators $X_i$ with the ones associated to the virtual copy $X'_i$, namely:
\begin{equation}
C'(X'_1,X'_2,X'_3;Y_1):=C_{\mathfrak{sl}_2}(X'_1,X'_2,X'_3)=(X'_2)^2+2(X'_1 X'_3+X'_3 X'_1)=(X'_2)^2+4 X'_1 X'_3+2 Y_1 X'_2 \, .
\label{eq:p}
\end{equation}
  After expanding this expression and reordering terms, taking into account that $Y_1=X_4$ is central, we get a fourth-order element that can be cast in the form:
\begin{equation}
C'(Y_1 X_1+Y_3^2/2, Y_1 X_2+Y_2 Y_3-Y_1/2, Y_1 X_3-Y_2^2/2; Y_1)=Y_1 C(X_1,X_2,X_3,Y_1,Y_2,Y_3)-3Y_1^2/4 \, ,
\label{eq:ex}
\end{equation}
from which, recalling the definitions \eqref{genrad}, we can extract the third-order Casimir \eqref{Casc}.

In this case, since we already know that the construction we used previously for the Levi factor is associated to the Racah algebra $R(3)$, we want to make advantage of the virtual copy in order to construct an abstract Racah algebra $R(3)$ working on the enveloping algebra of the whole Levi decomposable Lie algebra $\mathfrak{sl}_2 \niplus  \mathfrak{n}_{3,1}$. To this aim, taking into account that $Y_1$ is central, let us define the new generators: \begin{equation}
X''_i:=Y_1^{-1}X'_i \qquad i=1,2,3
\label{eq:ng}
\end{equation}
such that:
\begin{equation}
[Y_1, X_i'']=0 \, ,\qquad [X''_i, X''_j]=C_{ij}^kX''_k \, .
\label{eq:vcopres}
\end{equation}
The associated Casimir is:
\begin{equation}
C''(X''_1,X''_2,X''_3) =(X''_2)^2+2 (X''_1 X''_3+X''_3X''_1)= (X''_2)^2+4 X''_1 X''_3+2 X''_2 \, .
\label{cassec}
\end{equation}
We now apply our construction and introduce the intermediate Casimir invariants $C''_{\alpha}, C''_{\alpha \beta}$ and $C''_{123}$ associated to the Lie algebra \eqref{eq:vcopres}.  We know that in terms of these elements the quadratic algebra \eqref{1r}-\eqref{3r} is obtained. However, because of \eqref{eq:ng}, when we consider the expressions in terms of the generators $X_i'$ we obtain Casimir invariants with rational terms involving the central elements defined on each copy. This could be avoided by defining new polynomial Casimir invariants as we have done for $\mathfrak{n}_{5,5}$. However, we prefer to allow these rational terms involving central elements in the construction. So, what we want to do at this level is to re-express everything in terms of the original generators ($X_i$, $Y_i$) in order to show that it is possible to close a quadratic Racah algebra in terms of higher order polynomials defined in the enveloping algebra of the three copies of the Levi decomposable Lie algebra we are considering. To some extent, since it has been obtained from virtual copies of the levi factor, this quadratic algebra might be considered as a virtual copy of the quadratic Racah algebra.  In terms of three copies of the original generators ($X^{[\alpha]}_i$, $Y^{[\alpha]}_i$) for $i=1,2,3$ these elements explicitly read\footnote{We omitted  unessential constant terms.}:
\begin{equation}
C''_{\alpha} =X_2^{[\alpha]} + 4 X_1^{[\alpha]}  X_2^{[\alpha]}  + \bigl(X_2^{[\alpha]}\bigl)^2- 
2 \bigl(Y_1^{[\alpha]}\bigl)^{-1}X_2^{[\alpha]} \bigl(Y_2^{[\alpha]} \bigl)^2 + 2 \bigl(Y_1^{[\alpha]}\bigl)^{-1} X_2^{[\alpha]}Y_2^{[\alpha]}Y_3^{[\alpha]} + 2\bigl(Y_1^{[\alpha]}\bigl)^{-1}
 X_3^{[\alpha]}\bigl(Y_3^{[\alpha]}\bigl )^2
\end{equation}
\begin{align}
C''_{\alpha \beta} =&\,   4 X_1^{[\alpha]}  X_3^{[\alpha]} + 4 X_1^{[\alpha]}  X_3^{[\beta]}  + 4 X_1^{[\beta]}  X_3^{[\beta]}  + \bigl(X_2^{[\alpha]}  \bigl)^2 + 
2 X_2^{[\alpha]}  X_2^{[\beta]}  +\bigl(X_2^{[\beta]}  \bigl)^2 + 4 X_3^{[\alpha]}  X_1^{[\beta]} - \bigl(Y_1^{[\alpha]}\bigl)^{-1} Y_2^{[\alpha]}Y_3^{[\alpha]} \nonumber\\
&-\bigl(Y_1^{[\beta]}\bigl)^{-1} Y_2^{[\beta]}Y_3^{[\beta]} - 2\bigl(Y_1^{[\alpha]}\bigl)^{-1} X_1^{[\alpha]} \bigl(Y_2^{[\alpha]} \bigl)^2- 
2 \bigl(Y_1^{[\beta]}\bigl)^{-1}X_1^{[\alpha]} \bigl(Y_2^{[\beta]} \bigl)^2- 2 \bigl(Y_1^{[\beta]}\bigl)^{-1}X_1^{[\beta]}  \bigl(Y_2^{[\beta]}\bigl)^2 \nonumber\\
&+ 
2 \bigl(Y_1^{[\alpha]}\bigl)^{-1} X_2^{[\alpha]}  Y_2^{[\alpha]} Y_3^{[\alpha]}  + 2 \bigl(Y_1^{[\beta]}\bigl)^{-1} X_2^{[\alpha]}  Y_2^{[\beta]} Y_3^{[\beta]}   + 
2 \bigl(Y_1^{[\beta]}\bigl)^{-1} X_2^{[\beta]}  Y_2^{[\beta]} Y_3^{[\beta]}  + 2 \bigl(Y_1^{[\alpha]}\bigl)^{-1}X_3^{[\alpha]} \bigl(Y_3^{[\alpha]}\bigl)^{2}\nonumber \\
&+ 
2 \bigl(Y_1^{[\beta]}\bigl)^{-1} X_3^{[\alpha]} \bigl(Y_3^{[\beta]}\bigl)^{2}  + 2 \bigl(Y_1^{[\beta]}\bigl)^{-1}  X_3^{[\beta]} \bigl(Y_3^{[\beta]}\bigl)^{2} - 
2 \bigl(Y_1^{[\alpha]}\bigl)^{-1} \bigl(Y_2^{[\alpha]}\bigl)^{2}  X_1^{[\beta]} + 2 \bigl(Y_1^{[\alpha]}\bigl)^{-1} Y_2^{[\alpha]} Y_3^{\alpha} X_2^{\beta} \nonumber\\
&+ 
2 \bigl(Y_1^{[\alpha]}\bigl)^{-1}\bigl(Y_3^{[\alpha]}\bigl)^2  X_3^{[\beta]} - \bigl(Y_1^{[\alpha]}\bigl)^{-1}\bigl(Y_1^{[\beta]}\bigl)^{-1}\bigl(Y_2^{[\alpha]}\bigl)^2  \bigl(Y_3^{[\beta]}\bigl)^2+ 
2 \bigl(Y_1^{[\alpha]}\bigl)^{-1}\bigl(Y_1^{[\beta]}\bigl)^{-1}Y_2^{[\alpha]}  Y_3^{[\alpha]}  Y_2^{[\beta]} Y_3^{[\beta]} \nonumber \\
&- \bigl(Y_1^{[\alpha]}\bigl)^{-1}\bigl(Y_1^{[\beta]}\bigl)^{-1}\bigl(Y_3^{[\alpha]}\bigl)^2 \bigl(Y_2^{[\beta]}\bigl)^2
\end{align}
and we omit the expression of $C''_{123}$, which is anyhow related to these through the linear relation:
\begin{equation}
C''_{123}= C''_{12} +  C''_{23} +  C''_{13} - C''_1 - C''_2 - C''_3 \, .
\label{eq:c123}
\end{equation}
\noindent  These elements satisfy the commutation relations: $[C''_\alpha, C''_{12}]=[C''_\alpha, C''_{23}]=[C''_\alpha, C''_{13}]=0$, $[C''_{\alpha}, C''_{123}]=0$ and $[C''_{\alpha \beta}, C''_{123}]=0$. Moreover, once introduced the new generator:
\begin{equation}
C''_{1223}:= [C''_{12}, C''_{23}]= [C''_{23}, C''_{13}]=[C''_{13}, C''_{12}]\, ,
\label{eq:vcg}
\end{equation}
a direct computation shows that they close in a quadratic Racah algebra $R(3)$:
\begin{align}
&[C''_{12}, C''_{1223}]=8\bigl(C''_{23}C''_{12}-C''_{12}C''_{13}+(C''_2-C''_1)(C''_3-C''_{123})\bigl) \label{1rv}\\
&[C''_{23}, C''_{1223}]=8\bigl(C''_{13}C''_{23}-C''_{23}C''_{12}+(C''_3-C''_2)(C''_1-C''_{123})\bigl)\label{2rv}\\
&[C''_{13}, C''_{1223}]=8\bigl(C''_{12}C''_{13}-C''_{13}C''_{23}+(C''_1-C''_3)(C''_2-C''_{123})\bigl) \label{3rv}\, ,
\end{align}
where we have again $[C''_{12}, C''_{1223}]+[C''_{23}, C''_{1223}]+[C''_{13}, C''_{1223}]=0$.
\noindent We remark that the Casimir invariants $C''$ have been constructed by considering three copies of the generators $X''$. Each copy has been then re-expressed in terms of its associated copy related to the generators $X'$ and finally, taking into account \eqref{envalg}, in terms of the original generators $(X^{[\alpha]}_i, Y^{[\alpha]}_i)$, $i=1,2,3$.

\section{Conclusion}
\label{sec7}

In this paper we have investigated polynomial algebras obtained from intermediate Casimir invariants of Lie algebras, both simple and non-semisimple, in a purely algebraic setting, i.e. without relying to any realizations. We have proposed a systematic procedure and applied it to a variety of examples such as simple ($\mathfrak{sl}_{2}$, $\mathfrak{so}(1,3)$), nilpotent ($\mathfrak{n}_{5,5}, \mathfrak{n}_{6,1}$, $\mathfrak{n}_{6,19}$), solvable ($\mathfrak{s}_{6,160}$, $\mathfrak{s}_{6,183}$) and Levi decomposable  ($\mathfrak{sl}_2 \niplus 3\mathfrak{n}_{1,1}, \mathfrak{sl}_2 \niplus \mathfrak{n}_{3,1}$) Lie algebras. All the examples we have analyzed come from the classification of low dimensional (up to dimension six) indecomposable Lie algebras provided in \cite{sno14} (and we have followed the notation of this reference). 

The algebraic structures we have obtained present very different features, as they vary from Abelian algebras ($\mathfrak{n}_{5,5}$, $\mathfrak{n}_{6,19}$), algebras closed in terms of central elements on each copy of the initial Lie algebra ($\mathfrak{n}_{6,1}$), quadratic algebras ($\mathfrak{sl}_2$, $\mathfrak{so}(1,3)$, $\mathfrak{sl}_2 \niplus 3 \mathfrak{n}_{1,1}$) or even more complex structures involving higher order nested commutators ($\mathfrak{s}_{6,160}$, $\mathfrak{s}_{6,183}$). We have used the nilpotent Lie algebra $\mathfrak{n}_{5,5}$, for which we have obtained an Abelian structure from the intermediate Casimir invariants, as an illustrative example to show how the use of subalgebras can still provide a way to construct non-Abelian polynomial algebras. Finally, taking as a guiding example the Levi decomposable Lie algebra $\mathfrak{sl}_2 \niplus \mathfrak{n}_{3,1}$,  we have demonstrated how virtual copies of its Levi factor can be used to construct \textquotedblleft copies\textquotedblright of polynomial algebras, in fact, the quadratic Racah algebra $R(3)$ for our specific case.

The construction we have performed is completely abstract. However, once suitable realizations are considered, physical models associated to a given (chosen) algebraic Hamiltonian can arise. Notice that the search for realizations of Lie algebras is still ongoing and there exist classification results for low dimensional Lie algebras \cite{Popovych2003}. In the same spirit of \cite{CAMPOAMORSTURSBERG2021168378}, to build realizations one could for example rely on the coadjoint representation of $\mathfrak{g}$, that would in principle provide concrete differential operator realizations to play with at the level of the three copies. In any case, we might think of these algebraic structures as related to concrete physical models from many different perspectives once an \emph{algebraic Hamiltonian} has been identified. The explicit example we have sketched for the Lie algebra $\mathfrak{n}_{6,1}$ should at least give an idea of the potential applications. 
In fact, due to their purely algebraic nature, these results provide a scheme to obtain polynomial algebras underlying classes of Lie algebras which have found large applications both in classical and quantum mechanics.

\subsection*{Acknowledgements}

This work was supported by the Future Fellowship FT180100099 and Discovery Project DP190101529 from the Australian Research Council.

\subsection*{Data availability statement}

The computer code to reproduce the findings of this study is available upon request from the authors.

\addcontentsline{toc}{chapter}{Bibliography}
\bibliographystyle{utphys}
\bibliography{biblio}

\end{document}